\documentclass[
 reprint,
 amsmath,
 amssymb,
 aps,
 nofootinbib,
 superscriptaddress,
  prd
]{revtex4-2}
\usepackage[utf8]{inputenc}
\usepackage{placeins}
\usepackage{graphicx}
\usepackage{amsmath}
\usepackage{physics}
\usepackage{bm}
\usepackage{orcidlink}

\usepackage{dcolumn}
\newcolumntype{d}[1]{D{.}{.}{#1}}

\newcommand{\psb}{$\dot{P}_s^\mathrm{B}$ }
\newcommand{\bsurf}{$B_\mathrm{Surf}$ }

\newcommand{\apjl}{Astrophys. J.}
\newcommand{\mnras}{Mon. Not. R. Astron. Soc.}
\newcommand{\araa}{Annu. Rev. Astron. Astrophy.}
\newcommand{\aj}{Astron. J.}
\newcommand{\aap}{Astron. Astrophys.}
\newcommand{\sovast}{Sov. Astron.}

\newcommand{\apjs}{Astrophys. J. Suppl. Ser.}
\newcommand{\pasj}{Publ. Astron. Soc. Jap.}

\begin{document}

\title{Empirical Modeling of Magnetic Braking in Millisecond Pulsars to Measure the Local Dark Matter Density and Effects of Orbiting Satellite Galaxies}



\author{Thomas Donlon II \orcidlink{0000-0002-7746-8993}} \email{thomas.donlon@uah.edu}
\affiliation{Department of Physics and Astronomy, University of Alabama in Huntsville, 301 North Sparkman Drive, Huntsville, AL 35816, USA}

\author{Sukanya Chakrabarti \orcidlink{0000-0001-6711-8140}}
\affiliation{Department of Physics and Astronomy, University of Alabama in Huntsville, 301 North Sparkman Drive, Huntsville, AL 35816, USA}

\author{Sophia Vanderwaal\orcidlink{0009-0004-6864-1338}}
\affiliation{Department of Physics and Astronomy, University of Alabama in Huntsville, 301 North Sparkman Drive, Huntsville, AL 35816, USA}

\author{\;\\Lawrence M. Widrow \orcidlink{0000-0001-6211-8635}}
\affiliation{Department of Physics, Engineering Physics and Astronomy, Queen’s University, Kingston, ON K7L 3N6, Canada}

\author{Scott Ransom\orcidlink{0000-0001-5799-9714}}
\affiliation{National Radio Astronomy Observatory, 520 Edgemont Rd., Charlottesville, VA 22903, USA}

\author{Enrico Ramirez-Ruiz\orcidlink{0000-0003-2558-3102}}
\affiliation{Department of Astronomy and Astrophysics, University of California, Santa Cruz, CA 95064, USA}

\begin{abstract}
We present a novel method that enables us to estimate the acceleration of individual millisecond pulsars (MSPs) using only their spin period and its time derivative. 
For our binary MSP sample, we show that one can obtain an empirical calibration of the magnetic braking term that relies only on observed quantities. 
We find that such a model for magnetic braking is only valid for MSPs with small surface magnetic field strengths ($<3\times10^8$ G) and large characteristic ages ($>$ 5 Gyr). With this method we are able to effectively double the number of pulsars with line-of-sight acceleration measurements, from 27 to 53 sources. This expanded dataset leads to an updated measurement of the total density in the midplane, which we find to be $\rho_0$ = 0.108 $\pm$ 0.008 \textit{stat}. $\pm$ 0.011 \textit{sys} M$_\odot$/pc$^3$, and the first $>3\sigma$ measurement of the local dark matter density from direct acceleration measurements, which we calculate to be $\rho_{0,\mathrm{DM}}$ = 0.0098 $\pm$ 0.0025 \textit{stat.} $\pm$ 0.0003 \textit{sys}. M$_\odot$/pc$^3$ (0.37 $\pm$ 0.10 GeV/cm$^3$). This updated value for $\rho_{0,\mathrm{DM}}$ is in good agreement with literature values derived from kinematic estimates. 
The pulsar accelerations are very asymmetric above and below the disk; we show that the shape and size of this asymmetry can be largely explained by the north-south asymmetry of disk star counts and the offset in the Milky Way disk and halo centers of mass due to the Large Magellanic Cloud. \\\vspace{0.5cm}
\end{abstract}

\maketitle

\section{Introduction} \label{sec:intro}

For more than a century, astronomers have used the positions and velocities of stars in order to estimate accelerations produced by the gravitational field of the Milky Way \citep[MW,][]{BlandHawthornGerhard2016}. These studies typically rely on assumptions such as dynamical equilibrium and symmetry in order to relate the phase space information of stars to the underlying gravitational potential. 

Recent advancements in precision time-domain astronomy have allowed us to use precise time-series data to directly measure the accelerations of objects due to the Galaxy's gravity \citep{Chakrabarti2020}. These acceleration measurements, which are free of the typical assumptions that are made in kinematic studies, are particularly useful tools for assessing the fundamental properties and dynamical structure of our Galaxy \citep[for example,][]{Arora2024}. By mapping the acceleration field of the Galaxy, it should eventually be possible to determine the distribution of dark matter in the MW with fairly high accuracy. 

These new techniques now enable measurements of accelerations across different scales; especially interesting examples of this are black holes around visible stars \citep{Chakrabarti2023,El-Badry2023} and stars near the Galactic center \citep{Ghez1998,Ghez2005,Ghez2008,Gillessen2009,Genzel2010}. Both of these experience accelerations on the order 3 cm/s$^2$, which is many order of magnitude larger than that of Galactic accelerations (just 3$\times$10$^{-8}$ cm/s$^2$). Direct acceleration measurements now enable the new field of ``real-time'' Galactic dynamics in which we can now measure these very small accelerations that arise from the mass distribution of our Galaxy.

The only class of object that has been used to successfully measure Galactic accelerations up until now is binary millisecond pulsars \citep[MSPs,][]{Chakrabarti2021,Moran2023,Donlon2024}. However, it is expected that direct acceleration measurements will be enabled by other independent techniques in other ways; this should happen in the near future with eclipse timing \citep{Chakrabarti2022}, and by the end of the decade with ongoing extreme precision radial velocity observations \citep{Chakrabarti2020}. Binary MSPs have been a successful class of accelerometers because if one assumes that general relativity is correct (which it appears to be up to a very high level of precision, e.g. \citealt{WeisbergHuang2016}), then it is possible to exactly determine the Galactic acceleration for a binary system given its orbital parameters, the distance to the system, and its proper motion \citep[e.g.][]{DamourTaylor1991}. 

The restriction to only binary MSPs is problematic for two reasons. First, it is often difficult to measure the orbital period derivative of a system, which is very small. Similarly, it can be difficult to measure the masses of the objects in a binary pulsar system, which requires an independent measurement of a Shapiro delay for systems in circular orbits \citep[which is true for most MSPs,][]{CondonRansom2016}. Second, not all MSPs are in binary systems, which reduces the number of available acceleration measurements.

Both of these problems would be resolved if we were able to use the spin period information of a pulsar to measure an acceleration instead. This is possible in principle, and it is much easier to measure the spin information of pulsars; practically every MSP has a spin period derivative measurement. In practice, this is difficult because the emission of pulsars pumps away angular momentum, which slows the rotation of these objects over time. The exact spindown rate relies on the shape and strength of the pulsar's magnetic field, and there is not yet a satisfactory theoretical description for pulsar magnetic fields that can reproduce a given pulsar's spindown rate from first principles \citep{CondonRansom2016}. This makes it challenging to determine the exact contribution to the spindown of a MSP due to the underlying Galactic acceleration. 

Pulsar spin information has previously been used to obtain approximate accelerations, with the caveat that one must account for the substantial and unknown spindown rate, which is often larger than the underlying Galactic acceleration that one wishes to measure. For example, Phillips et al. \cite[][hereafter P21]{Phillips2021} statistically inferred the aggregate distribution of intrinsic spindown rates from a sample of MSPs while simultaneously fitting the Galactic acceleration near the Sun. However, that work was unable to measure the acceleration of any individual pulsar using its spin period information, or to obtain Galactic accelerations at other locations using this method. Similarly, Heflin \& Lieu \cite{HeflinLieu2021} show that the spin period derivative is not enough to constrain accelerations for individual pulsars because of intrinsic spindown from magnetic effects, although the addition of the second derivative of the spin period could provide a viable method for measuring the velocities and accelerations of pulsars. 

We present here for the first time a way to extract a reliable measurement of the Galactic acceleration using only the spin information of a MSP. This relies on an empirical model for the intrinsic magnetic spindown rates of pulsars, which is calibrated to observations of binary MSPs for which the spindown rates are known exactly. This doubles the number of pulsars that produce usable acceleration data. Additionally, the extended dataset permits us to obtain a more precise estimate of the dark matter density in the Galactic midplane than previous pulsar studies (by nearly a factor of 2). 

This new dataset also reveals a substantial asymmetry in the vertical acceleration profile of the Galactic disk. We illustrate and explain how this feature can be caused by a combination of disequilibrium features in the disk star positions, the offset in the MW disk and halo due to the effects of orbiting satellites.

\section{Data}

This work uses two overlapping sets of data. The first dataset (which we call the ``D24'' dataset) consists of the 25 binary MSPs from Donlon et al. \cite{Donlon2024} (hereafter D24), which have measured time derivatives of their orbital periods, parallaxes, and proper motions. These sources either have measured orbital eccentricity and masses of the pulsar and its companion, or an orbital period of more than 5 days, because relativistic effects are expected to be minimal for systems with long orbital periods. In addition to these sources, we also add three new pulsars for which timing solutions with all of the required parameters have recently been obtained: J1012-4235 \citep{Gautam2024}, J1518+4904 \citep{Ding2023,Tan2024}, and J0218+4232 \citep{Du2014,VerbiestLorimer2014,Tan2024}. We have removed J2043+1711 from this dataset, as it has been shown to have a large peculiar acceleration that is inconsistent with the Galactic potential, and is likely caused by some intrinsic effect and/or accelerations from orbital companions/nearby objects \citep{Donlon2024b}. 

After these adjustments, there are a total of 27 binary pulsars in the D24 dataset. It should be noted that the double-pulsar system J0737$-$3039A/B only provides a single acceleration measurement for the two pulsars, which share orbital parameters.

The second dataset (which we call the ``ATNF'' dataset) consists of all pulsars in the Australia Telescope National Facility (ATNF) Pulsar Catalogue \citep{atnf} that have a measured time derivative of the spin period, proper motion, and a distance derived from a parallax measurement. Pulsars that are associated with globular clusters were removed from the dataset, because the accelerations from a globular cluster can be orders of magnitude larger than the Galactic acceleration \citep{Phinney1993,Freire2001,Prager2017}. Additionally, we removed pulsars that are known to be interacting with their orbital companions, which can cause variations in the orbital and spin periods of those systems. There are 182 pulsars that satisfy these criteria. Note that one pulsar in the final dataset (PSR J2322-2650) has a planet, although we elect to keep it in the dataset because the planet is accounted for in the timing solution for that pulsar, i.e., for the spin-period variation \citep{Spiewak2018}.  However, the binary period-drift has not been measured, and earlier work on measuring Galactic accelerations from binary pulsars have not included this source \citep{Chakrabarti2021,Moran2023,Donlon2024}. It has one of the lowest radio luminosities measured to date, and thus far only permits a measurement of the spin period variation.

The data used and generated within this paper are publicly available at \url{https://github.com/thomasdonlon/Empirical_Model_MSP_Spindown_Accels}.

\section{Accelerations and Spindowns Due to Magnetic Braking} \label{sec:accel_spindown}

\begin{figure}
    \centering
    \includegraphics[width=\linewidth]{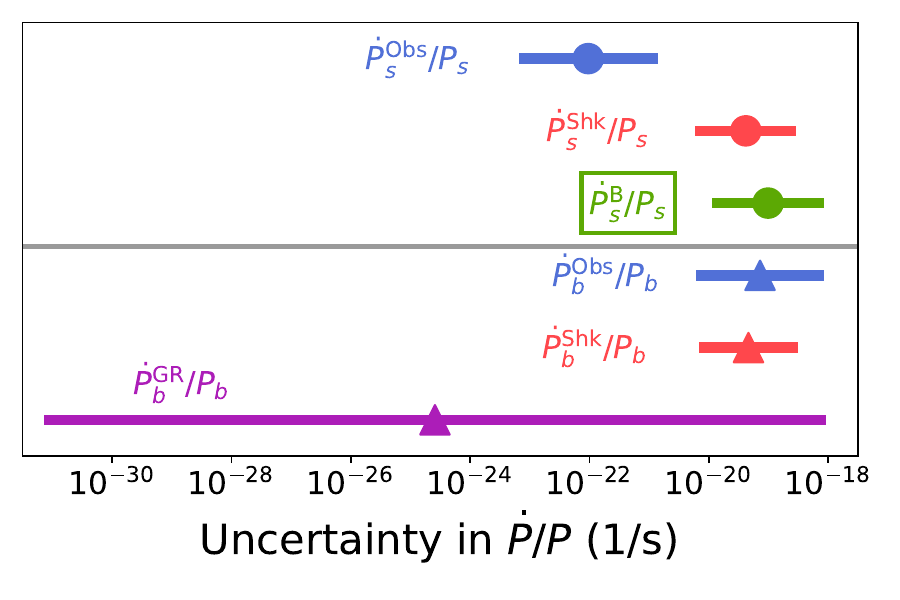}
    \caption{Uncertainties in the different components of spin and binary orbital period derivatives for all pulsars in the D24 sample. The central point indicates the mean uncertainty in each component, and the standard deviation of the uncertainty distribution is shown as lines on either side. Spin period data is shown as circles, while binary orbital period information is shown as triangles. The magnetic spindown term, $\dot{P}_s^\mathrm{B}/P_s$ (green, highlighted with a box), has a similar uncertainty to the Shklovskii terms and $\dot{P}_b^\mathrm{Obs}/P_b$, indicating that it is reasonable to construct an empirical model using this value. }
    \label{fig:uncertainties}
\end{figure}

In general, the line-of-sight velocity of an object leads to a Doppler shift in any periodic signal emitted by that object. A change in the velocity of that object over time, which is an acceleration, will result in a change in the Doppler shift of the periodic signal. This is expressed by the formula \begin{equation}
    \frac{\dot{P}}{P} = \frac{a_\mathrm{los}^\mathrm{tot}}{c},
\end{equation} where $P$ is the period of the emitted signal, $a_\mathrm{los}^\mathrm{tot}$ is the total line-of-sight acceleration of the object, and $c$ is the speed of light. Note that if the object is being accelerated by multiple sources, $a_\mathrm{los}^\mathrm{tot}$ and $\dot{P}$ will include contributions from each of these effects. 


Our goal in this section is to obtain the acceleration due to the gravity of the Milky Way ($a_\mathrm{los}$) using only the observed spin period ($P_s$) and its time derivative ($\dot{P}^\mathrm{Obs}_s$) for a given pulsar. However, we cannot simply use the measured value of $\dot{P}^\mathrm{Obs}_s$ for this, because there are several different effects that contribute changes in the observed spin period of the pulsar: \begin{equation} \label{eq:psdot}
    \dot{P}^\mathrm{Obs}_s = \dot{P}^\mathrm{Gal}_s + \dot{P}^\mathrm{Shk}_s + \dot{P}^\mathrm{B}_s,
\end{equation} where ``Obs'' is the observed change in the pulsar's spin period, ``Gal'' is the contribution from the pulsar accelerating due to the Galaxy's gravitational potential, ``Shk'' is the apparent change of the periodic signal due to the pulsar's proper motion on the sky and is known as the Shklovskii Effect \citep{Shklovskii1970}, and ``B'' is the magnetic braking\footnote{Note that ``magnetic braking'' has a different meaning in the context of binary stars (where surface magnetic fields lead to stellar winds, which carry away angular momentum); here we refer to the terminology of, for example, \cite{CondonRansom2016}. } of the pulsar due to its radiation pumping away angular momentum. 

The Shklovskii Effect is calculated as \begin{equation} \label{eq:shklovskii}
    \dot{P}^\mathrm{Shk} = \frac{P \mu^2 d}{c}.
\end{equation} This effect always has positive sign, and can often be large compared to the underlying Galactic acceleration term. 

If there were a straightforward way to calculate $P_s^\mathrm{B}$, then it would be simple to compute $\dot{P}^\mathrm{Gal}_s$ and therefore $a_\mathrm{los}$. Unfortunately -- crucially -- the magnetic braking term is not well understood, in that there is no theoretical model that can accurately determine \psb for a given pulsar from first principles. This is likely due to the magnetic field at the surface of realistic pulsars not being true dipoles, but instead being complex and dominated by high order multipoles, as has been shown to be the case in X-ray observations \citep{Riley2019,Kalapotharakos2021,Miller2021}. 

To further complicate things, previous attempts to measure \psb using measurements of $\dot{P}_s^\mathrm{Obs}$ have required estimating $\dot{P}^\mathrm{Gal}_s$ from kinematic models of the Galaxy's gravitational potential. These models are only constrained to the few tens of percent level if the Galaxy is assumed to be in equilibrium \citep{BlandHawthornGerhard2016}. However, common Galactic potential models could easily be incorrect by a factor of two or more in regions with substantial disequilibrium features, which have been shown to be prevalent in the MW disk \citep[e.g.][]{Widrow2012,Antoja2018,Donlon2024}.

We now describe a way to calculate $\dot{P}^\mathrm{Gal}_s$ rigorously for some pulsars, without assuming a model for the Galactic potential. Previously, accelerations have been calculated from the orbital period information of binary MSPs \citep[e.g.][]{Chakrabarti2021}. Changes in these orbital periods can be measured very precisely, and are subject to a different set of effects than the spin period; \begin{equation} \label{eq:pbdot}
    \dot{P}^\mathrm{Obs}_b = \dot{P}^\mathrm{Gal}_b + \dot{P}^\mathrm{Shk}_b + \dot{P}^\mathrm{GR}_b,
\end{equation} where the ``B'' term has been removed, and a ``GR'' term has been added. This GR term describes the relativistic decay of the binary orbit due to the emission of gravitational waves, and can be calculated as \begin{align*}
    \dot{P}_b^\textrm{GR} = -\frac{192\pi G^{5/3}}{5c^5}\left(\frac{P_b}{2\pi}\right)^{-5/3} \left(1-e^2\right)^{-7/2}
\end{align*}\begin{equation}\label{eq:pdot_gr}
    \times \left(1 + \frac{73}{24}e^2 + \frac{37}{96}e^4\right)\frac{m_p m_c}{\left(m_p + m_c\right)^{1/3}},
\end{equation} where $e$ is the orbital eccentricity of the binary, $m_p$ is the mass of the pulsar, and $m_c$ is the mass of its companion \citep{WeisbergHuang2016}. 

Once $\dot{P}^\mathrm{Gal}_b$ has been computed for a given binary pulsar, we can exactly calculate the magnetic braking term using the formula \begin{equation} \label{eq:psb_def}
    \dot{P}_s^\mathrm{B} = \dot{P}_s^\mathrm{Obs} - \frac{P_s}{P_b}
    \left( \dot{P}_b^\mathrm{Obs} - \dot{P}_b^\mathrm{GR}\right),
\end{equation} which is obtained by combining Equations \ref{eq:psdot} and \ref{eq:pbdot}, while recognizing that the relative changes in the spin period and orbital period of the pulsar must be equal, because they are both proportional to the Galactic acceleration:\begin{equation}
    \frac{a_\mathrm{los}}{c} = \frac{\dot{P}^\mathrm{Gal}_b}{P_b} = \frac{\dot{P}^\mathrm{Gal}_s}{P_s}. 
\end{equation} We note that the Shklovskii terms cancel out in Equation \ref{eq:psb_def}, so $\dot{P}_s^\mathrm{B}$ does not depend on the proper motion or distance of the pulsar. 

The relative uncertainties in all of these different terms are shown in Figure \ref{fig:uncertainties}. Note that because we are interested in accelerations (which are proportional to $\dot{P}/P$) and the scales of $\dot{P}_s$ and $\dot{P}_b$ are very different, we compare $\dot{P}/P$ rather than the value of $\dot{P}$ for these pulsars. Figure \ref{fig:uncertainties} shows that the uncertainty of $\dot{P}_s^\mathrm{B}/P_s$ is comparable to the other terms in Equations \ref{eq:psdot} and \ref{eq:pbdot}, indicating that we are able to calculate \psb with a high degree of accuracy. The uncertainty of the various $\dot{P}_s$ terms are similar to, or better than, their corresponding $\dot{P}_b$ terms. This suggests that we should be able to compute accelerations using the spin period information alone, so long as we are able to obtain a good estimate for $\dot{P}_s^\mathrm{B}$.

\section{An Empirical Model for Magnetic Braking} \label{sec:magnetic_braking_model}

\begin{figure*}
    \centering
    \includegraphics[width=\linewidth]{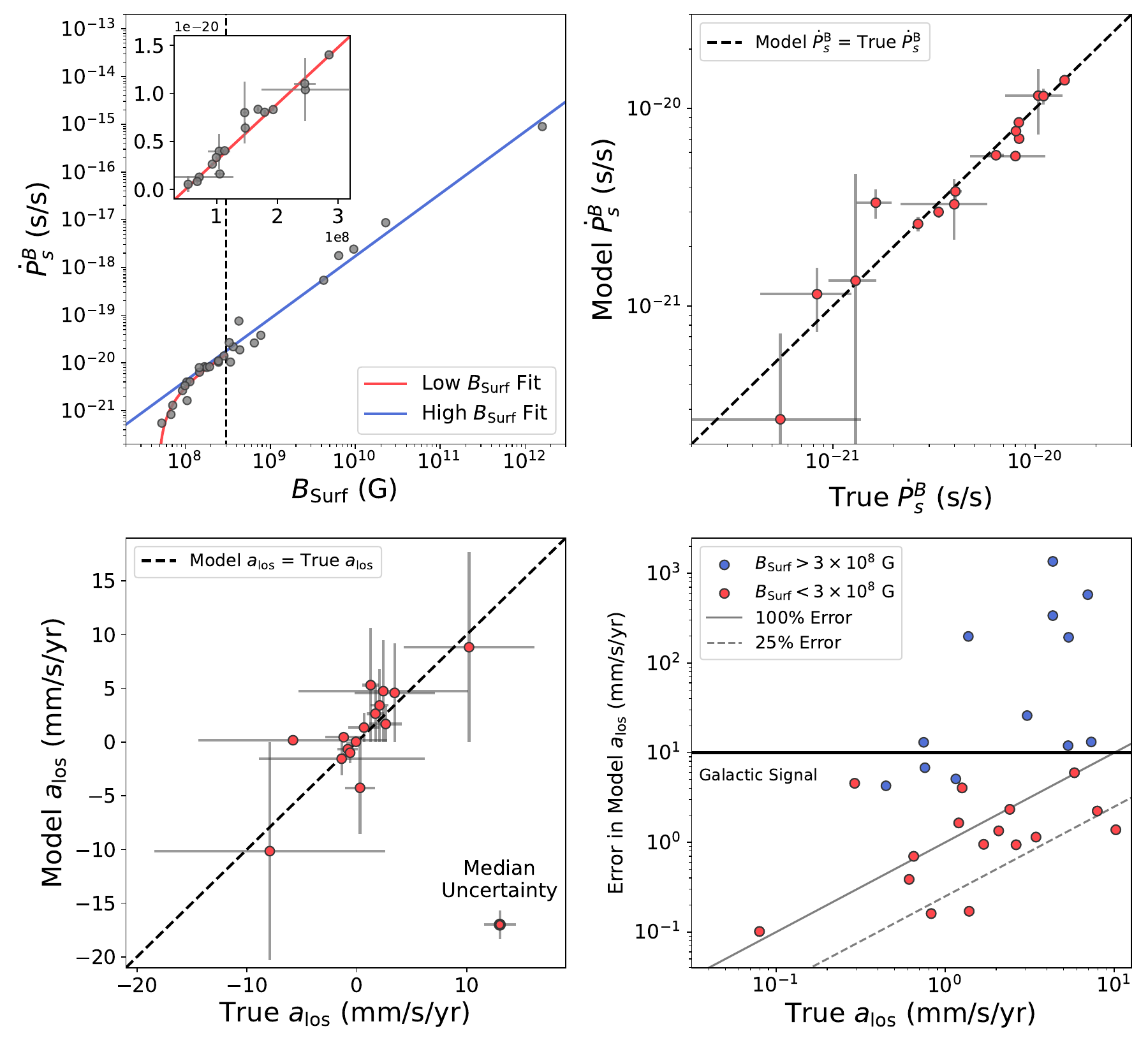}
    \caption{The performance of the empirical model for $\dot{P}_s^\mathrm{B}$ (spindown due to magnetic braking). The top left panel shows the empirical model for \psb given an inferred value of $B_\mathrm{Surf}$; overall the pulsar data is fit by a power law, but pulsars with \bsurf $<3\times10^8$ G are better fit by a linear model. The top right panel shows a comparison of the modeled and true values of $\dot{P}_s^\mathrm{B}$ for pulsars with \bsurf $<3\times10^8$ G, and the bottom left panel shows a comparison of the modeled accelerations and the accelerations calculated using binary orbital information for the same pulsars. Note that the median uncertainty of the accelerations from the empirical model and binary orbital period information are essentially identical. The bottom right panel shows the error in the line-of-sight acceleration for all pulsars in the D24 dataset; pulsars with lower values of \bsurf have smaller acceleration errors, indicating that our empirical model is only valid for these low-$B_\mathrm{Surf}$ pulsars. In general, the modeled values agree with the baseline values within the 1$\sigma$ uncertainties for almost all pulsars.}
    \label{fig:fit_and_errors}
\end{figure*}

One might expect there to be a relation between the rate of magnetic braking spindown for a given pulsar and the strength of that pulsar's magnetic field. This arises from the idea that the power radiated away by the rotation of the pulsar's magnetic moment scales with the strength of the magnetic moment. In addition, the length scale for any feedback torque exerted on the magnetized pulsar from its external magnetic field should extend out to the Alfven radius of the pulsar, which will be larger for a stronger magnetic field. The exact physics of these scenarios are complicated, predominantly because the true magnetic moments of pulsars are not actually dipoles. Nevertheless, the spindown rate of a pulsar is expected to depend somehow on the pulsar's magnetic field strength. 

If one assumes that the magnetic field of a pulsar is in-fact a dipole, then the minimum magnetic field strength at the surface of the pulsar is given as \begin{equation} \label{eq:bsurf_true}
    B_\mathrm{Surf} = \gamma \left(\dot{P}^B_s P_s\right)^{1/2},
\end{equation} where $\gamma$ = 3.2$\times$10$^{19}$ G s$^{-1/2}$ \citep{CondonRansom2016}. While this is not necessarily an accurate description of the true magnetic field strength at the surface of a given pulsar, it should be a valid estimate to within an order of magnitude. 

If the effect of Galactic acceleration is small compared to the spindown from magnetic braking ($\dot{P}_s^\mathrm{Gal} \ll \dot{P}_s^\mathrm{B}$), and the surface magnetic field strength is close to its minimum value, then this can be estimated as \begin{equation} \label{eq:bsurf_approx}
    B_\mathrm{Surf} \approx \gamma \left[\left(\dot{P}^\mathrm{Obs}_s - \dot{P}^\mathrm{Shk}_s\right) P_s\right]^{1/2}.
\end{equation} This quantity is useful because it relates easily-observed properties of a pulsar to a simple theoretical model of magnetic braking. The validity of this approximation is explored further in Appendix \ref{app:bsurf_approx}; we find that it is an appropriate approximation for this particular use case.

It is important to note that \bsurf is not a directly measured quantity, and should not be confused with the true magnetic field strength of the pulsar, which is unknown. Throughout the rest of this work, when we refer to a pulsar's ``(surface) magnetic field'', we mean the value of $B_\mathrm{Surf}$.

A quantity similar to \bsurf is characteristic age, which is defined as \begin{equation}
    \tau = \frac{P_s}{2\dot{P}^\mathrm{B}_s} \approx \frac{P_s}{2(\dot{P}^\mathrm{Obs}_s - \dot{P}^\mathrm{Shk}_s)}.
\end{equation} For non-recycled pulsars, this is an estimate of the time that a pulsar has been radiating away energy. However, characteristic age is not indicative of the actual time since a MSP underwent a supernova \citep{Tauris2012}, because the recycling process substantially moves a pulsar across the $P_s$-$\dot{P}_s$ diagram \citep{CondonRansom2016}. Despite this, characteristic age presumably still contains information about the properties of a MSP.

\subsection{Model for Known Values of $\dot{P}_s^B$}

Since we know the true value of $\dot{P}^B_s$ for the D24 pulsars from Section \ref{sec:accel_spindown}\footnote{Except for B1534+12, which was excluded from this section because it has a negative $\dot{P}_s^B$, which is non-physical.}, we can fit a model to the measured values of $\dot{P}^B_s$ and the observed and/or derived quantities of those pulsars. Because we expect the value of \psb to scale with $B_\mathrm{Surf}$, we first attempt to model $\dot{P}^B_s$ as a power law of \bsurf according to the relation \begin{equation} \label{eq:power_law}
    \dot{P}^B_s \approx \alpha B_\mathrm{Surf}^\lambda,
\end{equation} where the best-fit values for the power law model are $\log_{10}(\alpha/\mathrm{G})=-30.8\pm0.6$ and $\lambda=1.31\pm0.07$. This model is shown as blue line in the top left panel of Figure \ref{fig:fit_and_errors}. 

Note that in the top left panel of Figure \ref{fig:fit_and_errors}, the observed $\dot{P}_s^B$ data appears to fall below the power law fit from Equation \ref{eq:power_law} for low values of $B_\mathrm{Surf}$. Below 3$\times10^8$ G, the data is better modeled by the linear relation \begin{equation} \label{eq:linear_law}
    \dot{P}^B_s = m B_\mathrm{Surf} + b,
\end{equation} where the best-fit values for the linear model are $m=5.9\pm0.1\times10^{-29}$ G$^{-1}$ and $b=-2.8\pm0.2\times10^{-21}$. This model for \psb is shown as a red line in the top left and inset panels of Figure \ref{fig:fit_and_errors}. This model provides a way of estimating \psb that only depends on the directly and easily observable values $\dot{P}_s$ and $P_s$, rather than binary orbital information.

As there appears to be significant statistical noise in the data, we also show the results of this fit while including an intrinsic scatter term in Appendix \ref{app:int_err}. However, the addition of an intrinsic scatter term does not change the values of the fit parameters outside of their 1-$\sigma$ uncertainties.

The relative errors in the fit for \psb and the corresponding acceleration inferred from that value of the magnetic spindown are shown in the remaining panels of Figure \ref{fig:fit_and_errors}, respectively. The linear model does a good job of fitting these quantities for the D24 dataset, as the modeled values agree with the true values within the 1$\sigma$ uncertainties for almost all pulsars in the sample. However, pulsars with \bsurf $>$ 3$\times$10$^8$ G are not shown on these panels, because their errors are larger than the borders of the bottom left panel. Additionally, in the bottom right panel of Figure \ref{fig:fit_and_errors}, it is clear that the residuals are much smaller for the pulsars with $B_\mathrm{Surf} < 3\times10^8$ G than they are for the pulsars with large surface magnetic field strengths. From this, we conclude our model is only viable for low-$B_\mathrm{Surf}$ MSPs. Note that we cannot reliably estimate $\dot{P}_s^B$ for the well-timed Hulse-Taylor pulsar (B1913+16) or the double pulsar (J0737$-$3039), because they are only partially recycled.

We point out that, for the low-\bsurf pulsars, the median uncertainty of the modeled accelerations (vertical error bars in the bottom left panel of Figure \ref{fig:fit_and_errors}) is essentially identical to the median uncertainty of the pulsar accelerations calculated using the binary orbital period information (horizontal error bars). Because the uncertainties of the accelerations from the spin period model are similar to the accelerations from binary orbit information, we claim that -- on average -- this model produces accelerations with the same level of precision as the binary orbit data. This is not an outrageous result, because many of the D24 acceleration measurements from binary pulsar timing data have fairly large uncertainties. To summarize, for MSPs with small \bsurf and large characteristic age, the average estimated acceleration is currently ``as good'' as the average acceleration from binary orbital period information.

\subsection{Assessing Acceleration Bias}

\begin{figure}
    \centering
    \includegraphics[width=\linewidth]{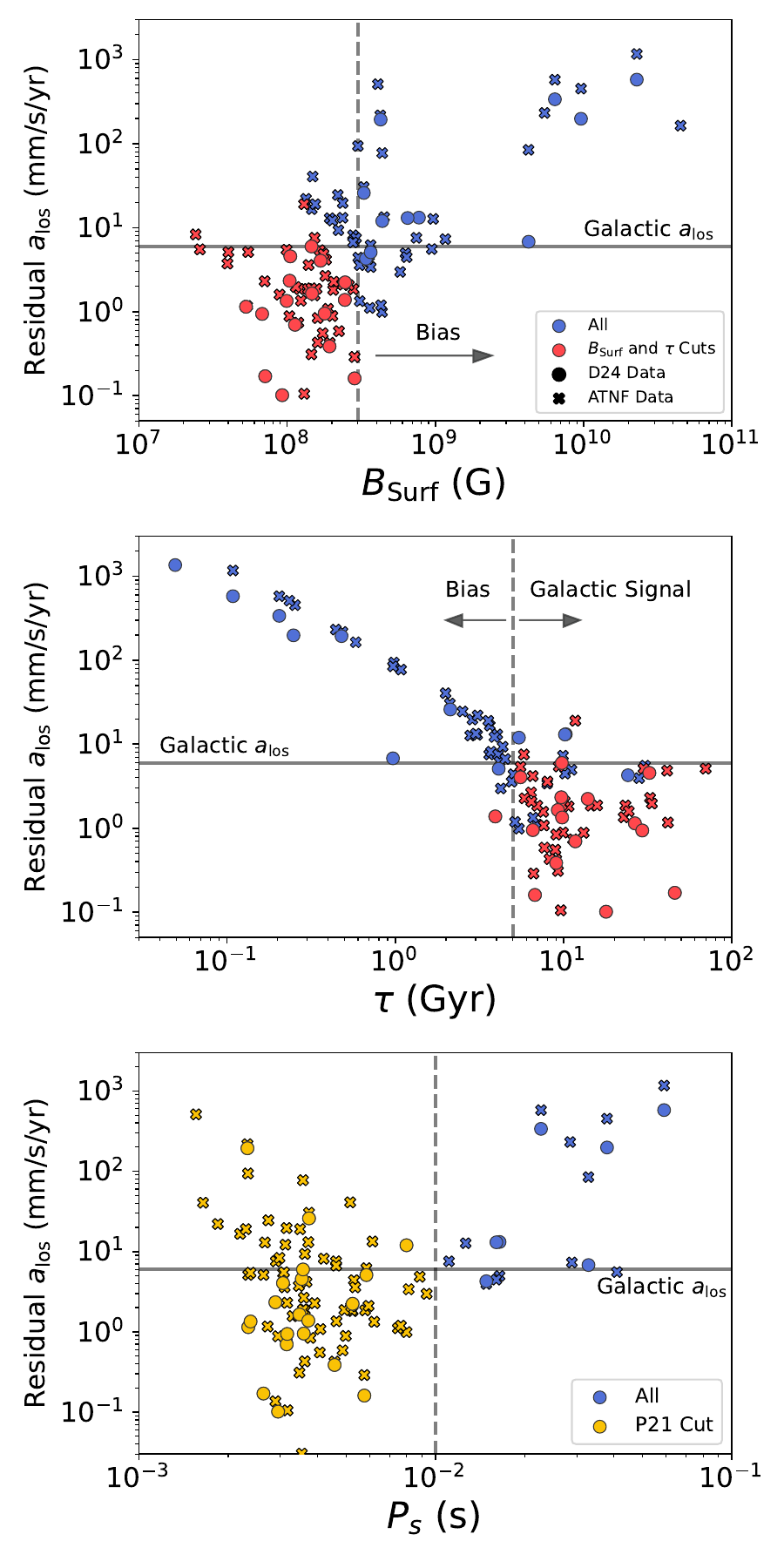}
    \caption{Residuals between modeled line-of-sight accelerations and kinematic predictions for Galactic accelerations. The approximate magnitude of the Galactic acceleration is shown as a horizontal line. In the top two panels, the red points show pulsars with low \bsurf strengths and large characteristic ages; these objects have ``intrinsic'' accelerations that are smaller than the Galactic signal, indicating that they can be used to measure Galactic accelerations. Acceleration bias is worse in pulsars with large \bsurf and small characteristic age. In the bottom panel, the yellow points show pulsars with $P_s<0.01$ ms, which has been used in previous studies \citep{Phillips2021} to remove partially-recycled pulsars. The low-$P_s$ selection contains many pulsars with significant acceleration bias, emphasizing the need for the cuts in \bsurf and $\tau$. }
    \label{fig:char_age_bias}
\end{figure}

Now that we have a model for $\dot{P}_s^B$ that relies only on measured pulsar spin period observables, we compute $\dot{P}_s^B$ for the entire ATNF dataset. We can estimate the accuracy of the model by comparing the accelerations we calculate using the magnetic spindown model and the accelerations that are predicted by a kinematic potential model at the location of each pulsar. The kinematic model that we use in this case is the \verb!gala! potential model \verb!MilkyWayPotential2022! \citep{Gala}. This model was fit to recent observations of the MW's properties, and as a result is a reasonable time-static approximation of the Galactic potential. 

Figure \ref{fig:char_age_bias} shows the residual acceleration plotted against the surface magnetic field strength and characteristic age of each pulsar. It is clear that our empirical model only performs well for pulsars with $B_\mathrm{Surf} < 3\times10^8$ G and $\tau > 5$ Gyr (shown as vertical dashed lines). There are 40 pulsars in the ATNF dataset that satisfy these criteria, 14 of which are in the D24 dataset. This brings the total number of pulsars up to 53, essentially doubling the number of pulsars that can be used to compute reliable accelerations compared to D24. 

\subsection{Alternative Analysis of Intrinsic Spindown}

Previously, P21 statistically inferred the distribution of intrinsic spindown rates from a population of pulsars along with the acceleration at the location of the Sun. However, they were unable to use the spin periods of individual pulsars to measure an acceleration at other points in space. One reason for this is because many pulsars in their sample have large intrinsic spindowns, which dramatically changes the inferred acceleration for these sources; P21 had no way to calculate the intrinsic spindown of each pulsar to access the underlying acceleration information. 

P21 selected pulsars based on their spin period, only using pulsars with $P_s<10$ ms for their study. Their reasoning for this cut was to remove partially recycled pulsars, which could potentially have atypical values for their magnetic braking spindown because they presumably have a different composition than MSPs. 

We show the effect of using the P21 cut in $P_s$ to select our pulsar sample in the bottom panel of Figure \ref{fig:char_age_bias}. A substantial number of these pulsars have significantly biased inferred accelerations that will prevent the underlying Galactic acceleration from being accurately measured. This makes it clear that a cut in $P_s$ alone is not sufficient to select pulsars that can be used with our empirical model.

\section{Physical Interpretation of the Empirical Model} \label{app:phys_interp}

\begin{figure}
    \centering
    \includegraphics[width=\linewidth]{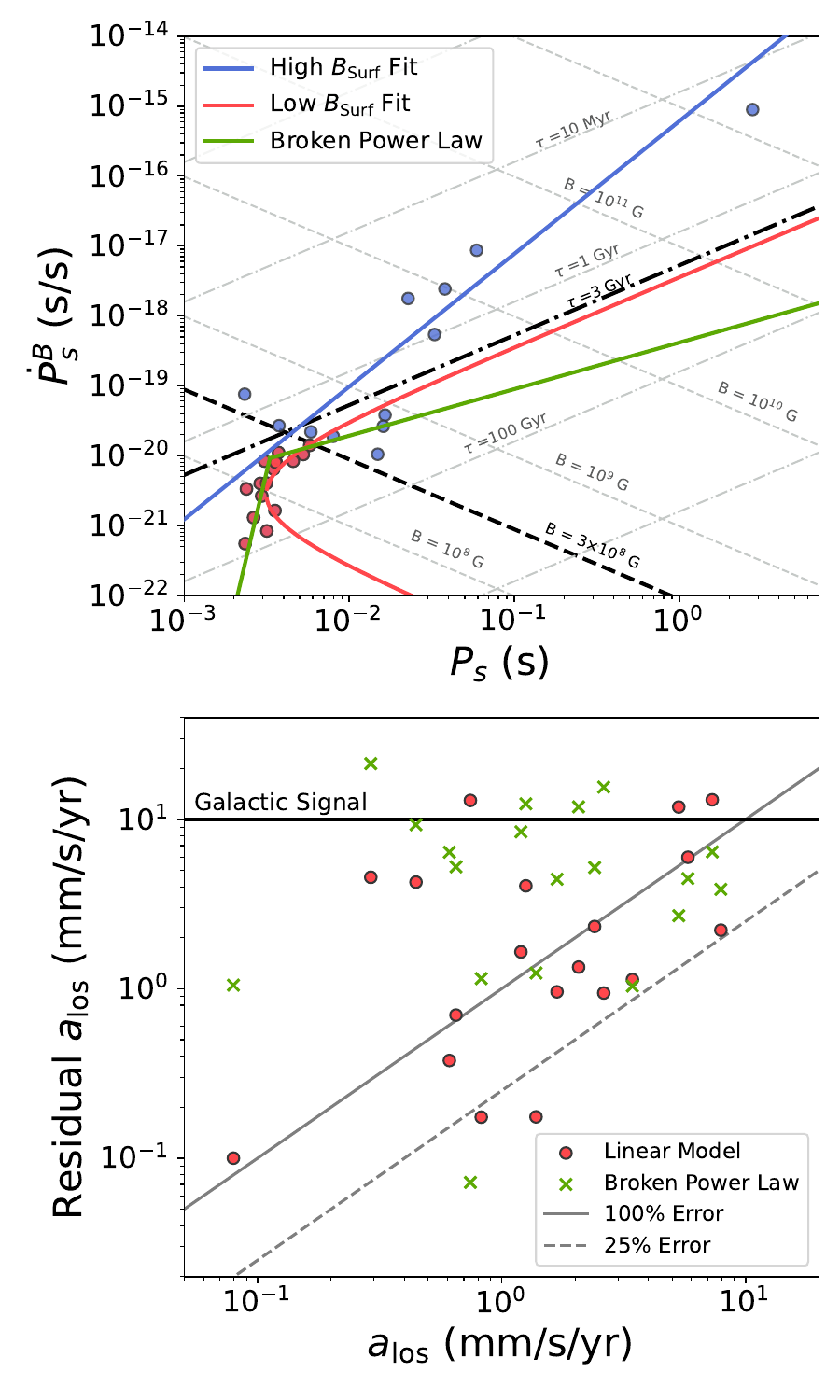}
    \caption{An alternative broken power-law model for computing $\dot{P}_s^\mathrm{B}$ from only $P_s$ (green). Although the broken power-law model appears to match the pulsar data better in $\dot{P}_s^B$ vs. $P_s$ space, it performs worse than the linear model in Equation \ref{eq:linear_law} (shown as the red line, which appears as a sideways parabola in this projection). The power-law fit of equation \ref{eq:power_law} is shown as a blue line. MSPs with very rapid periods ($P_s <$ 33 ms) appear to have much smaller values of $\dot{P}_s^\mathrm{B}$ than pulsars with slower spin periods, potentially indicating that the low-\bsurf MSPs have different magnetic braking behavior than pulsars with larger values of $B_\mathrm{Surf}$. }
    \label{fig:power_law_model}
\end{figure}

It is potentially difficult to intuitively understand the underlying physical meaning of the empirical model used in this work, because \bsurf is a derived property rather than a measured quantity. Further, there is no guarantee that \bsurf is an accurate estimate of the true magnetic field strength for a given pulsar. On one hand, this is acceptable because, by virtue of being an empirical model, we are really only concerned with the performance of the model rather than its physical interpretation. On the other hand, one wonders whether there is some clearer interpretation of the empirical model that can shed light on the inner workings of MSPs. 

We can convert the empirical models from Equations \ref{eq:power_law} \ref{eq:linear_law} into a functional form that express $\dot{P}^B_s$ as a function of only $P_s$, as well as the fit parameters and constants. Solving for $\dot{P}^B_s$ produces the following relationship for the power law model: \begin{equation}
    \dot{P}^B_s = \left[ \alpha \gamma^\lambda P_s^{\lambda/2}\right]^{2/(2-\lambda)},
\end{equation} which for the fit values of $\alpha$ and $\lambda$ result in a relationship of $\dot{P}^B_s\propto P_s^{1.90}\approx P_s^2$. Although this functional form looks like a braking index relation, it should not be interpreted as one; this is the relationship of $\dot{P}^B_s$ and $P_s$ for a population of pulsars at a given time rather than a single pulsar over time.

Solving for $\dot{P}^B_s$ in the linear model produces the relationship: \begin{equation}
    \dot{P}^B_s = \frac{1}{2}\left[ m^2 \gamma^2 P_s + 2b \pm m\gamma\sqrt{P_s}\sqrt{m^2\gamma^2P_s + 4b} \right], 
\end{equation} which is a sideways parabola in $\dot{P}^B_s-P_s$ space. 

In order to better understand this relationship, we show the low-\bsurf pulsars in the D24 dataset in $P_s-\dot{P}^B_s$ space in the top panel of Figure \ref{fig:power_law_model}. We note that the shape of the points in this figure appear to not be a parabola, but would possibly be better fit by a broken power law in $P_s-\dot{P}^B_s$ space. Motivated by this, we define a new model, described by \begin{equation}
    \dot{P}_s^B = \begin{cases} 
      k_1 P_s^{\lambda_1} & P_s < P_0, \\
     k_2 P_s^{\lambda_2} & P_s \geq P_0,
   \end{cases}
\end{equation} where the optimized parameters are $k_1$ = 6.4$\times10^{4}$, $\lambda_1$ = 10, $k_2=4.146\times10^{-19}$, $\lambda_2=2/3$, and $P_0$ = 3.3$\times10^{-3}$ s.\footnote{The units of $k_i$ are such that $\dot{P}_s^B$ has units of s/s given an input of $P_s$ in s.} While this model is more aesthetically pleasing in that it does not involve an intermediary derived quantity such as $B_{\rm Surf}$, we show in the bottom panel of Figure \ref{fig:power_law_model} that it does not perform as well at recovering accelerations as the linear model described by Equation \ref{eq:linear_law}. 

The physical interpretation of the relationship between spindown rates and $P_s$ in pulsars with large characteristic age is not immediately clear. The spindown rates of these pulsars appear to increase dramatically from very small spin periods until a spin period of $\sim$0.003 s, at which point the spindown rate for a given $P_s$ roughly follows a line of constant characteristic age. We speculate that this could indicate some (lack of a) magnetic process occurring inside very rapidly spinning pulsars that causes them to have small intrinsic spindown rates; such a process might only be active in pulsars on one side of the $P_s\sim$ 0.003 s threshold. Future work could reveal some physical explanation for this observed relationship between intrinsic spindown rate and spin period. 


\subsection{On Pulsar Companions}

\begin{figure}
    \centering
    \includegraphics[width=\linewidth]{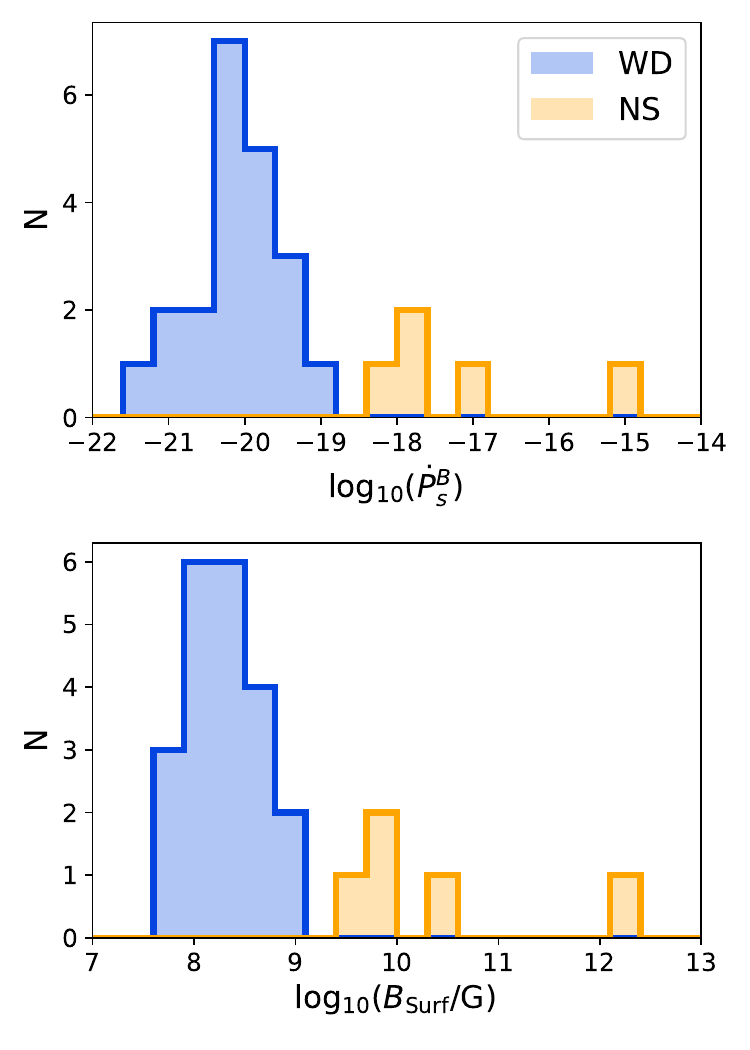}
    \caption{Magnetic properties of the binary pulsar sample, split into whether their orbital companion is a white dwarf or a neutron star. The surface magnetic field (and therefore the amount of magnetic braking) is much larger for pulsars with neutron star companions than for pulsars with white dwarf companions; as a result, our empirical method is only applicable to pulsars with white dwarf companions. }
    \label{fig:magnetic_hist}
\end{figure}

Figure \ref{fig:magnetic_hist} shows a breakdown of the magnetic properties of the binary pulsar sample, split up into whether the orbital companion is a neutron star or a white dwarf. It is clear that pulsars with neutron star companions have much larger rates of spindown due to magnetic braking (and correspondingly higher surface magnetic fields). This indicates that our empirical model for magnetic braking will only be valid for pulsars with white dwarf companions. 

The questions remains: why are the spindown rates for neutron star binaries so large compared to pulsar-white dwarf binaries? One possible reason for this is due to the magnetic field of the companion, which can induce twisting of the magnetic field lines and dissipation of the pulsar's spin energy as additional electromagnetic radiation \citep{HansenLyutikov2001,Lai2012}. The energy dissipated via this process is roughly \begin{equation}
    \dot{E}_\mathrm{diss} = 1.7\times10^{44} \left(\frac{B_\mathrm{comp}}{10^{13}\mathrm{ G}}\right)^2 \left(\frac{a}{\mathrm{30 km}}\right)^{-7} \mathrm{ erg/s}.
\end{equation} Neutron star companions have substantial magnetic fields, which could cause $\dot{E}_\mathrm{diss}$ to be large for neutron star--pulsar binaries, while a white dwarf companion would not lead to this effect due to its much weaker magnetic field.

However, this explanation seems unlikely; the shortest-period neutron star-neutron star binaries in our sample have periods of about 0.1 days, which corresponds to a separation of roughly 10$^6$ km. With this length scale, even if the companion had a particularly strong magnetic field of $B\sim10^{15}$ G, the dissipated power is only on the order of $10^{16}$ erg/s, whereas the power radiated due to dipole radiation is on the order of 10$^{35}$ erg/s. Due to the extremely strong scaling as the separation of the neutron stars becomes small, it appears that this effect is only relevant when the objects are close to merger. 

Some other process must be responsible for this discrepancy, although we are unable to isolate it in the context of this work. It seems reasonable that the difference between pulsars with white dwarf vs. neutron star companions may be related to their evolution and the spin up process, as all neutron star-pulsar binaries appear to be only partially recycled \citep{Kremer2024}. We speculate that perhaps the evolutionary formation channels of the two populations can lead to these differences. White dwarfs typically have smaller mass than neutron stars; it stands to reason that during the evolution of the binary system, if the pulsar accretes more mass off of its companion, that companion will wind up as a white dwarf rather than a neutron star.  The additional accreted mass during spin-up could lead to the pulsar having more angular momentum and its magnetic field being screened by additional surface mass in the white dwarf case compared to pulsars with neutron star companions. However, the evolutionary scenarios for neutron star binaries are particularly complicated \citep[e.g.][]{Tauris2017}, so more work is required to determine whether this is actually the case.

\section{Local Acceleration Gradient}

\subsection{Pulsar Data}

We now move away from our discussion of magnetic braking and compact objects, and begin using the new acceleration measurements to infer properties of the MW. 

The primary phenomenon that will be discussed below is the vertical acceleration gradient near the Sun. This feature was initially discussed by Chakrabarti et al. \cite{Chakrabarti2021} (hereafter C21), who pointed out that in simulations, interactions between the MW and its orbiting satellites led to substantial disequilibrium features in the vertical acceleration profile that could be observed using direct acceleration measurements \citep{Chakrabarti2020}. The asymmetry in the vertical acceleration profile was then measured using pulsar timing data by D24, who found that the observed asymmetry agreed well with a simulation of a Sagittarius (Sgr) dwarf galaxy-like satellite interacting with a MW-mass disk. With the expanded pulsar acceleration dataset from this work, we now have a wealth of new data available that enables us to measure this acceleration asymmetry in even more detail. 

The vertical component of the line-of-sight acceleration can be written as a power series: \begin{align}
    a_{\mathrm{los},z}(z) &= \sum_{i=0} f_i z^i.
\end{align} It is useful to split this up into the even terms, which are symmetric about $z=0$, and the odd terms, which are asymmetric: \begin{equation}
    a_{\mathrm{los},z}(z) = \sum_{i=\mathrm{even}} f_i z^i + \sum_{i=\mathrm{odd}} f_i z^i.
\end{equation} We now define a quantity that measures the asymmetry of the vertical acceleration profile, which can be expressed using the odd terms of the above power series:\begin{align} \label{eq:delta_alosz_def}
    \Delta a_{\mathrm{los},z}(z) &= a_{\mathrm{los},z}(+z) - a_{\mathrm{los},z}(-z) \\
    &= 2\sum_{i=\mathrm{odd}} f_i z^i \nonumber
\end{align} Truncating this series at the second nonzero term, we obtain an approximation for $\Delta a_{\mathrm{los},z}(z)$: \begin{equation} \label{eq:delta_alosz}
    \Delta a_{\mathrm{los},z}(z) \approx 2f_1z + 2f_3z^3.
\end{equation} 

However, we do not directly measure $a_{\mathrm{los},z}$ from the data, we measure $a_\mathrm{los}$; this makes it less straightforward to determine $\Delta a_{\mathrm{los},z}(z)$ from the pulsar data. We can approximately obtain $a_{\mathrm{los},z}$ by subtracting off the predicted line-of-sight acceleration (computed using the Gala \textbf{MilkyWayPotential2022} model) and the observed radial bias in pulsar accelerations (see Section III of \citealt{Donlon2024} and Appendix A of \citealt{Donlon2024b}) from the observed $a_\mathrm{los}$ values. It should be noted that this will also remove part of the $a_{\mathrm{los},z}$ component; however, because the potential model is symmetric about $z=0$, subtracting off the acceleration due to the smooth potential will only remove the symmetric part of the $a_{\mathrm{los},z}$ distribution. This conveniently leaves us with the asymmetric part of the $a_{\mathrm{los},z}$ distribution, which is all that is needed to compute $\Delta a_{\mathrm{los},z}(z)$. 

The optimized parameters are $f_1 = -0.5$ $\pm$ 0.2 mm/s/yr/kpc and $f_3 = 0.3$ $\pm$ 0.5 mm/s/yr/kpc$^3$, which results in an observed value of $\Delta a_{\mathrm{los},z}(1\;\mathrm{kpc})$ = -0.4 mm/s/yr/kpc. Even though the $f_3$ term is not well constrained by the data, we find that including it produces more reasonable behavior of $\Delta a_{\mathrm{los},z}(z)$ at moderate-to-large $z$ heights. In the following subsections, we will discuss two potential sources for this observed asymmetry and determine their relative strengths.

\subsection{Disk Waves}

\begin{figure}
    \centering
    \includegraphics[width=\linewidth]{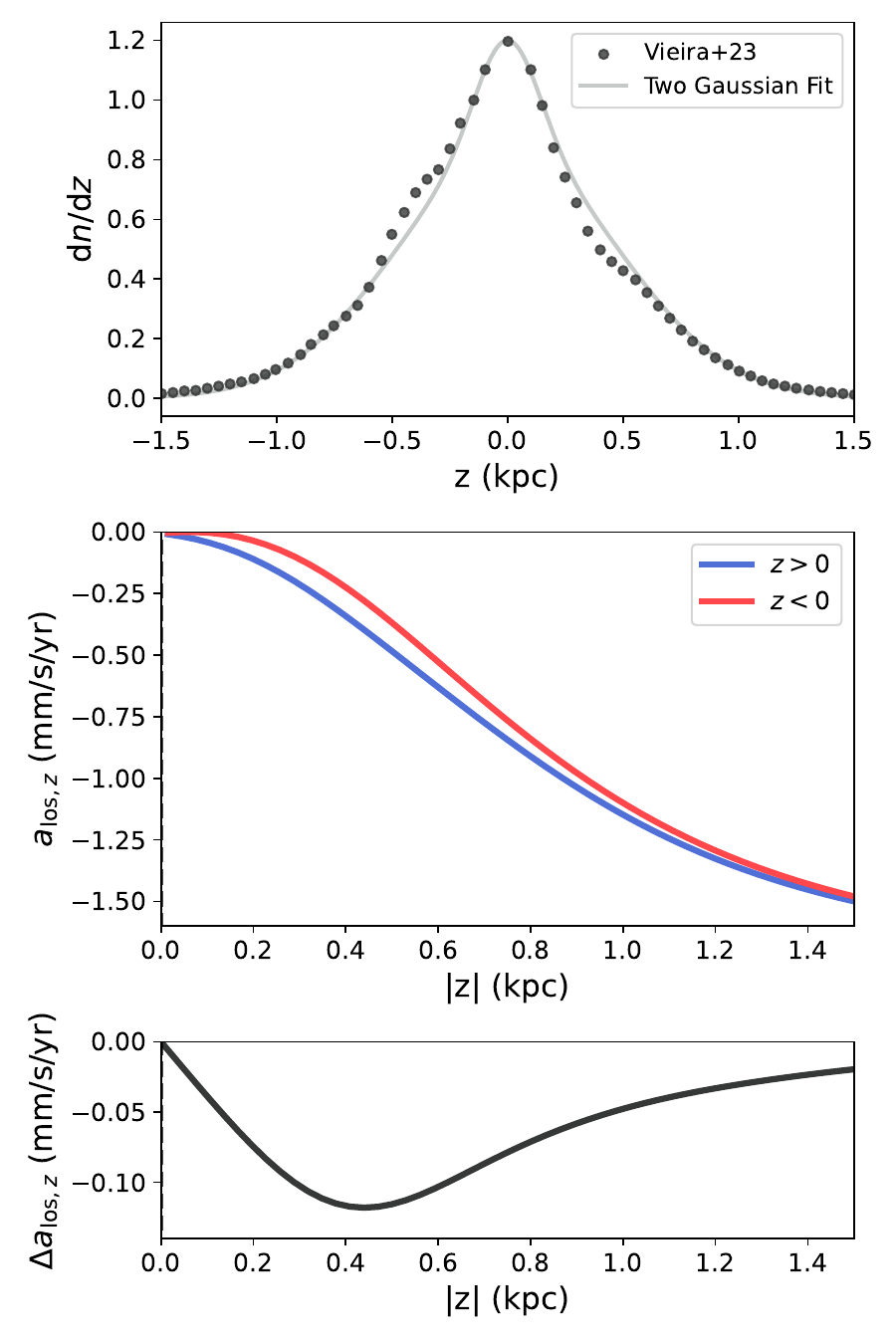}
    \caption{\textit{Top:} Star counts in the MW disk as a function of vertical position \citep[datapoints from][]{Vieira2023}. A prominent over/under-density can be seen at $z=\pm0.5$ kpc, which is related to vertical waves in the disk. A fit consisting of two Gaussian distributions centered on $z=0$ is shown in light grey, which is symmetric about the midplane. \textit{Middle:} The vertical acceleration profile of the Vieira et al. \cite{Vieira2023} data, which is asymmetric above and below the disk due to the perturbation shown in the top panel. The strength of the observed line-of-sight accelerations are larger (more negative) above the disk than they are below the disk. \textit{Bottom:} Difference between line-of-sight acceleration above and below the midplane for different vertical positions. Negative $\Delta a_{\mathrm{los},z}$ indicates that the magnitude of the line-of-sight acceleration at a given distance from the midplane is larger (more negative) above the midplane than below it. }
    \label{fig:disk_asym_alos}
\end{figure}

The MW disk has an asymmetry in star counts above and below the plane \citep{Widrow2012,YannyGardner2013,BennettBovy2019}. This asymmetry is related to waves in the disk, which are thought to be caused (at least in part) by interactions between the MW disk and orbiting satellites \citep[e.g.][]{Xu2015}. This density asymmetry could easily lead to a vertical acceleration gradient; near the Sun, there are fewer stars above the Galactic midplane than below it, causing the acceleration field to be stronger above the midplane than below it. 

The number density of disk stars as a function of vertical position \citep[taken from][]{Vieira2023} is shown as grey points in the top panel of Figure \ref{fig:disk_asym_alos}. The solid gray line in that panel shows a fit of two Gaussian distributions centered on $z=0$ to the star count data, which is guaranteed to be symmetric about $z=0$ in order to illustrate the magnitude of the north-south asymmetry in star counts. Although the vertical disk wave structure has been shown to extend beyond a kpc from the midplane \citep{BennettBovy2021}, in terms of star counts (and therefore mass) most of the deviation from symmetry occurs around $z=\pm0.5$ kpc. 

The vertical acceleration profile of the MW disk can be estimated by placing a series of point masses above and below the midplane, \begin{equation}
    a_{\mathrm{los},z}(z) = C\sum_i \frac{Gn(z_i)}{(z - z_i)^2}
\end{equation} where $z_i$ are the vertical positions of the Vieira et al. \cite{Vieira2023} datapoints, and $n(z_i)$ is the integral of the corresponding number density. The normalization factor $C$ converts between star counts and stellar mass; it is set so that $a_{\mathrm{los},z}$(1.5 kpc) = -1.5 mm/s/yr. This normalization is equal to the observed acceleration at that distance, and is in agreement with the theoretical disk model accelerations at that distance from \textit{Gala} and \textit{Galpy} models.  

The vertical acceleration profile, split up into the part above the plane (blue) and below the plane (red) is shown in the middle panel of Figure \ref{fig:disk_asym_alos}. It is clear that $a_{\mathrm{los}z}$ is weaker below the plane than it is above the plane. This leads to a nonzero $\Delta a_{\mathrm{los}z}$, which is shown in the bottom panel of Figure \ref{fig:disk_asym_alos}; it has a maximum value of roughly 0.11 mm/s/yr, and is positive below the plane and negative above the plane. However, this effect is only relevant at $|z|\sim$ 0.5 kpc, whereas the observed value of $\Delta a_{\mathrm{los}z}$ is large beyond $|z|\sim$ 1 kpc. As such, the vertical density asymmetry of the disk is not large enough on its own to explain the observed value of $\Delta a_{\mathrm{los}z}$.

\subsection{Disk/Halo Offset} \label{sec:disk_halo_offset}

\begin{figure}
    \centering
    \includegraphics[width=\linewidth]{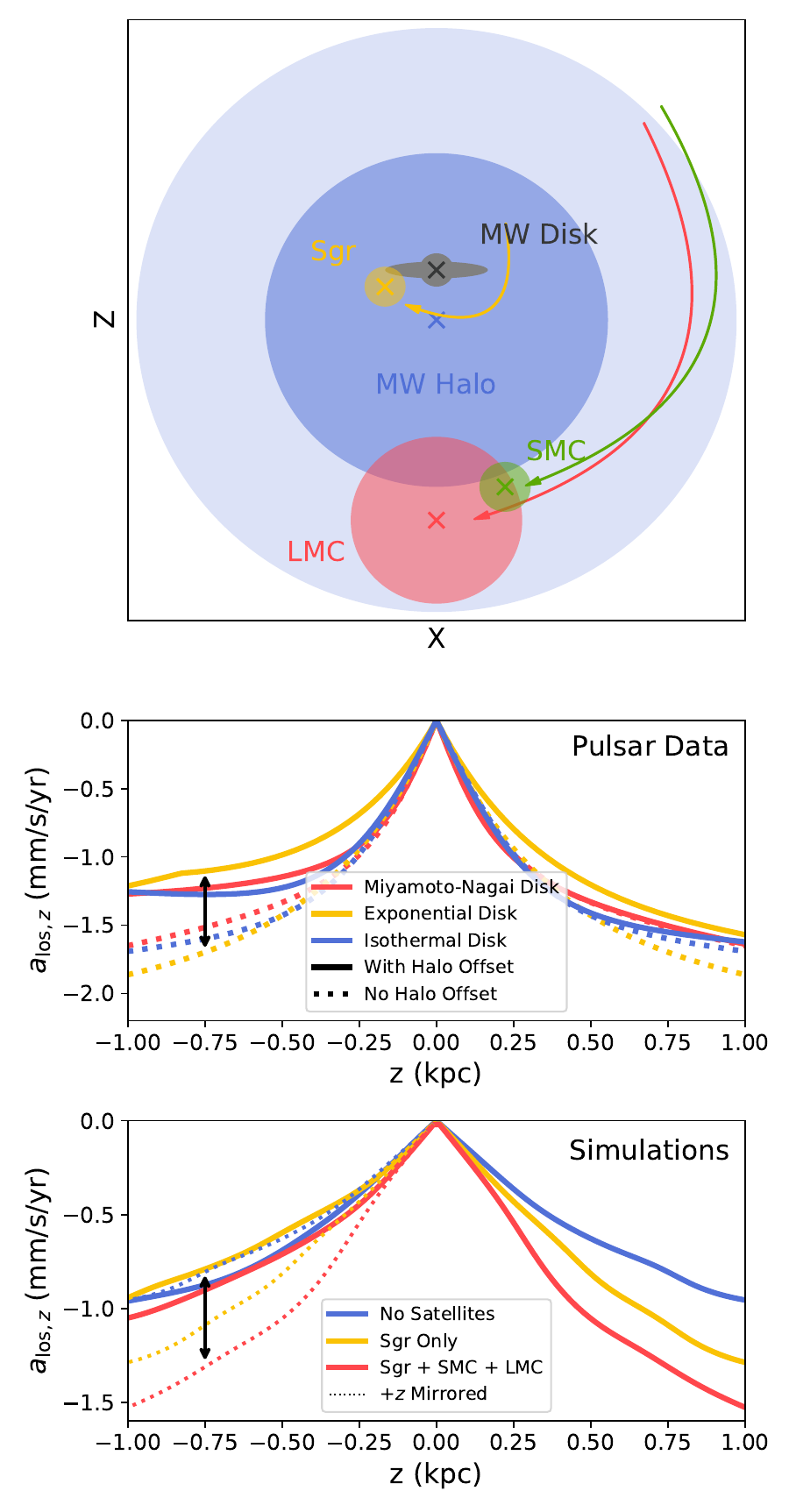}
    \caption{\textit{Top:} A simplified schematic of the MW halo and several orbiting satellite galaxies (not to scale). The approximate past paths of the satellites are traced by colored arrows. Centers of mass for each object is indicated by an ``x''. The gravitational influence of the orbiting satellites has dragged the MW disk center of mass away from the halo center of mass. \textit{Middle:} Vertical acceleration profile from the pulsar data. Two models (Miyamoto-Nagai disk and isothermal disk) are fit to the observed data; when an offset between the halo and disk components of the potential are included, we observe vertical accelerations that are smaller for negative $z$ values than corresponding positive values of $z$. \textit{Bottom:} Vertical acceleration profiles for simulations with a varying number of orbiting MW satellite analogues. Simulations with satellites exhibit the same asymmetry in $a_\mathrm{los}$, suggesting it is caused by an offset between the MW halo and disk centers of mass due to interactions with satellite galaxies. }
    \label{fig:alosz}
\end{figure}

Since the observed local acceleration gradient is too large to be caused by only the known vertical density asymmetry in the disk, we are led to examine another effect that can generate a vertical acceleration gradient. The most prominent interaction with the MW at the present day is the infall of the Large and Small Magellanic Clouds (LMC, SMC) which is substantially impacting the distribution of matter in the MW's halo \citep{Erkal2019,Garavito-Camargo2021,Vasiliev2021,PetersenPenarrubia2021,Vasiliev2023}. Additionally, the interaction between the MW and Sgr significantly perturbs the MW mass distribution (\cite{Laporte2019}, but see \cite{BennettBovy2021,Hunt2022}); however, this interaction mainly drives disequilibrium in the MW disk rather than the outer halo, so its effects are largely subsumed in our treatment of the disk waves. 

Although the acceleration on the Solar neighborhood due to the LMC is small, the interaction produces a large-scale tidal field that has shifted the MW disk center of mass away from the MW halo's center of mass. This is illustrated in the top panel of Figure \ref{fig:alosz}, where the inner halo has been shifted towards the LMC relative to the MW disk. This is a very simplified picture -- the inner halo is significantly deformed due to the passage of the LMC, and the LMC has its own halo that contributes a substantial amount of mass to the MW -- but it is nevertheless suitable to estimate the effect that the disk/halo offset has on the local acceleration gradient. 

In order to quantify the offset between the disk and halo components, we optimize several potential models to the observed pulsar data, but allow the halo to be shifted relative to the disk along the $z$ axis by an arbitrary amount (for specifics regarding the potential models and the fitting procedure, see Section \ref{sec:density}). The offset between the halo and disk was determined to be roughly 0.7 kpc towards the Galactic South; an offset of less than a few kpc is consistent with the distance between the disk and inner halo centers of mass predicted by simulations of the MW/LMC system \citep{Garavito-Camargo2021}. 

The effect of this offset on the vertical acceleration profile is shown in the middle panel of Figure \ref{fig:alosz}. Although the exact amount and shape of the acceleration asymmetry depends on the choice of disk potential, the value of $\Delta a_{\mathrm{los},z}$ is roughly -0.5 mm/s/yr at a height of 1 kpc. The asymmetry due to an offset in the disk and halo centers of mass is also present in simulations of the MW and interacting satellites \citep{Chakrabarti2019}, shown in the bottom panel of Figure \ref{fig:alosz}. When no satellites are included in the simulation, the effect disappears. When only a Sgr-like satellite is added, an asymmetry is produced, although it is smaller than that observed in the MW. When a Sgr analog plus analogs of the Magellanic Clouds are added to the simulation, we observe an asymmetry in the vertical acceleration profile that is similar to the observed MW value. 

We tried fitting the halo offset with and without including the vertical disk density asymmetry in the accelerations of the pulsar data, and found that it did not significantly change the results of the fit compared to the reported uncertainties; this is presumably because the disk density asymmetry is largest around $|z|=0.5$ kpc, and the halo offset effect is largest further from the midplane (between 1 and 3 kpc depending on the choice of potential model). In other words, the acceleration profiles generated by the two effects have different shapes and are sensitive to different regions of space.

Note that we have effectively included a ``full model'' of the accelerations imparted by the Sgr interaction at low $|z|$. Any effects Sgr has on MW accelerations near the Sun are either part of the disk waves (which are known exactly, because we use the observed disk density distribution), or orbital effects (in essence, the disk/halo offset, and the Sgr dwarf is included in our simulations). There will also be a direct acceleration in the direction of the dwarf remnant, but that is negligibly small at the Solar position.

\subsection{Comparison to the Observed Gradient}

\begin{figure}
    \centering
    \includegraphics[width=\linewidth]{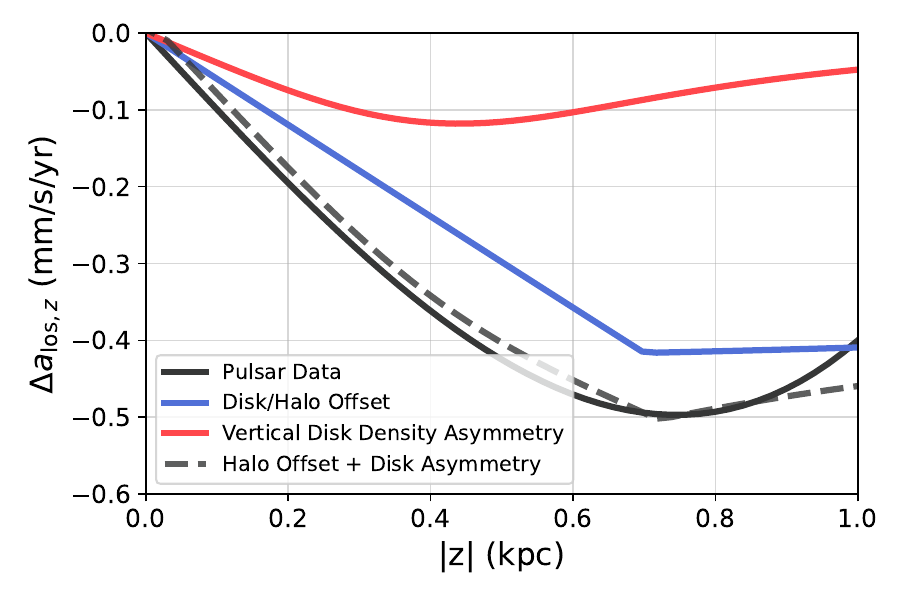}
    \caption{Difference between vertical acceleration above and below the midplane (Eq. \ref{eq:delta_alosz_def}), broken down into individual components. The solid black line shows the truncated power series fit to the pulsar data (Eq. \ref{eq:delta_alosz}), which indicates the observed gradient in line-of-sight acceleration. The red line is the contribution due to vertical disk density asymmetry, while the blue line is the contribution due to the fit offset between the disk and halo centers of mass; the dashed gray line shows the sum of these two effects, which has the same shape as the observed acceleration gradient from pulsar data, although the observed data has a larger magnitude. This could indicate the presence of an additional acceleration mechanism that we were not able to identify, or it may simply be due to uncertainty in the observed pulsar data. }
    \label{fig:accel_gradient}
\end{figure}

Figure \ref{fig:accel_gradient} shows the observed vertical acceleration gradient from the pulsar data, along with the estimates of the acceleration gradients that are produced by the vertical disk density asymmetry and the disk/halo offset due to the LMC. The sum of these two effects has roughly the same shape and magnitude as the observed vertical acceleration gradient. 

The contribution from the disk/halo offset dominates at large $z$ heights, where it is 5-10x larger than the contribution from the disk density asymmetry. The effect of the disk density asymmetry is more relevant within 0.5 kpc of the midplane, although it is never larger than the effect of the disk/halo offset.

\section{The Oort Limit and Local Dark Matter Density} \label{sec:density}

\begin{table*}[]
    \centering
    \begin{tabular}{lrrrrr}
    \hline
        Parameter & Isothermal Disk & \;\; Exponential Disk & \;\; Miyamoto-Nagai Disk & \;\; Gala & \;\; Galpy \\ \hline \hline
        $\sigma$ (\textit{km/s}) & 20.9 $\pm$ 0.5 & -- & -- & -- & -- \\
        Disk Scale Length (\textit{kpc}) & -- & -- & 3.1 $\pm$ 0.8 & -- & -- \\
        Disk Scale Height (\textit{kpc}) & 0.32 $\pm$ 0.04 & 0.21 $\pm$ 0.02 & 0.28 $\pm$ 0.03 & -- & -- \\
        $\log_{10}(M_\mathrm{Disk}/M_\odot)$ & -- & -- & 10.9 $\pm$ 0.3 & -- & -- \\
        $\xi$ (\textit{mm/s/yr}) & 1.6 & 1.6 & 1.6 & -- & -- \\
        $z_0$ (\textit{kpc}) & 0.69 $\pm$ 0.08 & 0.68 $\pm$ 0.03 & 0.7 $\pm$ 0.1 & -- & -- \\
        $\log_{10}(M_\mathrm{NFW}/M_\odot)$ & 11.96 $\pm$ 0.20 & 11.99 $\pm$ 0.07 & 11.99 $\pm$ 0.07 & -- & -- \\
        $\rho_0$ (\textit{M$_\odot$/pc$^3$}) & 0.089 $\pm$ 0.006 & 0.130 $\pm$ 0.009 & 0.103 $\pm$ 0.010 & -- & -- \\
        $\rho_{0,\mathrm{DM}}$ (\textit{M$_\odot$/pc$^3$}) & 0.009 $\pm$ 0.004 & 0.010 $\pm$ 0.002 & 0.010 $\pm$ 0.002 & -- & -- \\ \hline
        $\Delta$AIC & 1.0 & 0.0 & 2.2 & 65.0 & 129.4 \\ \hline \hline
          
    \end{tabular}
    \caption{Optimized parameters and AIC for MW potential models that fit to the D24+ATNF pulsar dataset. The AIC of the \textit{Gala} \textbf{MilkyWayPotential2022} and \textit{Galpy} \textbf{MWPotential2014} kinematic models are also provided for comparison. The fit disk and halo masses are slightly larger than in the kinematic models.}
    \label{tab:model_params}
\end{table*}

\subsection{Fitting Procedure} \label{sec:fitting}


To infer the gravitational potential (as well as the local density) from pulsar accelerations, we begin with a parametric model for the potential ($\Phi$). We use this to calculate the heliocentric acceleration by taking the directional derivative of the potential in the Galactic rest frame and subtracting off the acceleration at the position of the Sun: \begin{equation}
    a_\mathrm{los}^\mathrm{model}(\mathbf{x}_i) = -\left[ \nabla \Phi(\mathbf{x}_i) - \nabla \Phi(\mathbf{x}_\odot) \right] \cdot \hat{\mathbf{d}}_i,
\end{equation} where $\hat{\mathbf{d}}_i$ is the unit line of sight vector from the Sun to the $i$th pulsar. 

The log likelihood function is then given by \begin{align}
    \mathcal{L} =& \sum_i^n\frac{(a_\mathrm{i,los} - a_\mathrm{i,los}^\mathrm{model})^2}{2\sigma_{a_\mathrm{i,los}}^2}  \\ 
    &+ \sum_i^n\ln\left[1 + (a_\mathrm{i,los} - a_\mathrm{i,los}^\mathrm{model})^2/\xi^2\right] + n\ln(\pi\xi),  \nonumber
\end{align} and uncertainties in these fits were estimated via jackknife resampling. 

This likelihood function includes an additional noise term characterized by the parameter $\xi$, which can account for unknown sources of uncertainty in the measurements or intrinsic scatter in the model. Moran et al. \cite{Moran2023} found what appeared to be additional random contributions to the accelerations of individual pulsars; see that paper for discussion of some possible sources of stochastic accelerations. Our choice of a Lorentzian rather than Gaussian form for the noise term is motivated by the observation that these random accelerations appear to have more extended tails than a Gaussian distribution \citep{Moran2023}. Across all models, the fit value of $\xi$ was about 1.6 mm/s/yr. 

We only use the 48 out of 53 pulsars in the D24+ATNF dataset that are located within 3 kpc of the Sun for these fits, because distant pulsars have large uncertainties in distance that can lead to large errors in the inferred acceleration. Additionally, our relatively simple models are expected to only be a reasonable description of the gravitational potential within a few kpc of the Sun.

\subsection{Discussion of $\alpha$ Models}

\begin{figure}
    \centering
    \includegraphics[width=\linewidth]{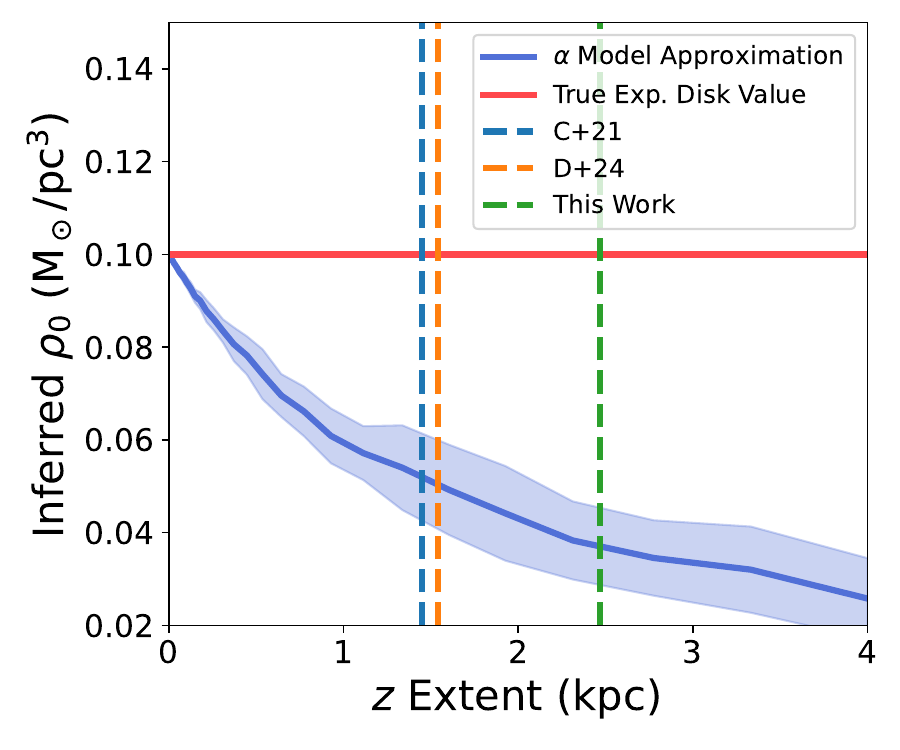}
    \caption{Midplane density is underestimated by $\alpha$ potential models (solid blue line with 1$\sigma$ uncertainty region), compared to realistic disk models where the density falls off as distance from the midplane increases (red line). In the case of an exponential disk potential with a scale height of 0.3 kpc, the $\alpha$ models will underestimate the midplane density by 40\% to 70\% within the vertical extents probed by the pulsar data (vertical dashed lines). This explains why previous pulsar-based measurements of the dark matter midplane density were so small compared to kinematic measurements. The $\alpha$ models are only valid at a 10\% level within a few hundred pc of the midplane. }
    \label{fig:exp_vs_alpha}
\end{figure}

C21 and D24 previously characterized pulsar accelerations using additively separable models for the potential, with the form $\Phi(R,z) = \Psi(R) + \zeta(z)$. These so-called $\alpha$ models have a vertical component made up of a truncated Taylor series; \begin{equation}
    \zeta(z) = \frac{\alpha_1}{2}z^{2} + \frac{\alpha_2}{3}z^3
\end{equation} and the radial component is \begin{equation}
    \Psi(R) = V_\mathrm{LSR}^2 \ln(R/R_\odot).
\end{equation} 

This set of $\alpha$ models is useful since they are easily related to the Oort limit -- the volume density in the Galactic midplane -- through the Poisson Equation \begin{equation} \label{eq:poisson}
    4\pi G \rho_0 = \alpha_1.
\end{equation} 
These models were intended to fit the local acceleration data only (within $\sim$ 1 kpc of the Sun), and are not suitable for use over the larger dynamic range that is now available for our current dataset.

We re-fit the various $\alpha$ models used in D24 to the extended pulsar dataset. These models all produced extremely low values for the Oort Limit, with a mean and standard deviation across the different models of 0.038 $\pm$ 0.013 M$_\odot$/pc$^3$. For comparison, the volume density of baryonic material in the Galactic midplane has been found to be $\rho_\mathrm{bary} = 0.84\pm0.013$ M$_\odot$/pc$^3$ \citep{McKee2015}. The $\alpha$ models imply a dark matter density of -0.046 $\pm$ 0.018 M$_\odot$/pc$^3$, which is unphysical and would be inconsistent at a $>2\sigma$ level with this baryonic budget. 

This issue stems from the fact that for these models, $\rho(z) = \rho_0$, and the global potential cannot have a constant density at all heights. While this approximation is suitable very close to the Galactic midplane, it breaks down at the few kpc heights above the disk that are now spanned by the pulsar data. The $\alpha$ models essentially average over the density in the probed region, and because the mean density between $\pm$1 kpc from the Galactic plane will be much lower than the density at the midplane, these models severely underestimate the midplane density. 

Figure \ref{fig:exp_vs_alpha} shows how the $\alpha$ models underestimate the density of a more realistic disk model, as a function of the vertical height probed by the data. The ``correct'' disk potential was taken to be an exponential disk with a central density of 0.1 M$_\odot$/pc$^{3}$ and a scale height of 0.3 kpc. We then generated 50 points, sampled uniformly within 1.5 kpc in the $X$ and $Y$ directions from the Sun, but varying the vertical extent of the samples in $Z$. An $\alpha$ model was then fit to the accelerations at each of the sampled points, the corresponding midplane density was calculated, and this process was repeated 100 times for each value of $z$ extent. Figure \ref{fig:exp_vs_alpha} shows that the $\alpha$ approximation is appropriate only for very small vertical extents (below a few hundred pc from the midplane). For the extents that are probed by the pulsar data, $\alpha$ models can underestimate the Oort limit by as much as 40\% to 70\%. This explains why the dark matter midplane densities measured by C21 and D24 were small compared to kinematic values. 

Note that if one assumes a ``correct'' Oort limit of 0.1 M$_\odot$/pc$^3$ \citep{McKee2015}, then the value of the Oort Limit measured by C21 (0.08 M$_\odot$/pc$^3$) would be underestimated by 20\%, and the value measured by D24 (0.062 M$_\odot$/pc$^3$) would be underestimated by 40\%. As such, the underestimation of the $\alpha$ models shown in Figure \ref{fig:exp_vs_alpha} is probably somewhat exaggerated (this is easily explained by the real MW disk not being well fit by a single exponential density profile). Regardless, it is clear that $\alpha$ models will systematically underestimate the Oort limit compared to more realistic density models where the density falls off with distance from the midplane.

\subsection{Improved Disk Models}

Since the $\alpha$ models are problematic, we seek a different method of calculating the Oort limit from the pulsar data. We use three physically-motivated models for the MW disk, where the density decreases as distance from the midplane increases. These more-realistic models could not be constrained from the available binary period pulsar data alone \citep{Donlon2024}, illustrating the importance of the additional spin period pulsar data produced in this work.

The first of these models is an isothermal disk potential \citep{BinneyTremaine2008}, which has a vertical component of the potential given by \begin{equation}
     \zeta(z) = \sigma^2\ln\left[\cosh\left(\frac{z}{h}\right)\right],
\end{equation} where $\sigma$ is the vertical velocity dispersion of the disk and $h$ is the scale height. We use a flat rotation curve for the radial component of the potential.

The second model uses an exponential density profile for the vertical component of the disk, \begin{equation} \label{eq:exp_density}
     \rho(z) = \rho_0 \mathrm{e}^{- |z| / h_z}, 
\end{equation} where $h_z$ is the scale height of the disk, and $\rho_0$ is the midplane density of the disk. The vertical component of the acceleration is then \begin{equation}
    a_z(z) = -4\pi G \rho_0 h_z \left(1 - e^{-|z|/h_z}\right)\mathrm{sgn}(z),
\end{equation} while the radial component of the acceleration is calculated assuming a flat rotation curve. 

The final disk potential model is a Miyamoto-Nagai disk \citep{MiyamotoNagai1975},  defined as \begin{equation}
    \Phi(R,z) = -\frac{GM_\mathrm{tot}}{\sqrt{R^2 + (a + \sqrt{z^2 + b^2})^2}},
\end{equation} where $M_\mathrm{tot}$ is the total mass of the disk, and $a$ and $b$ are the scale length and height of the disk, respectively.

In addition to the disk models, we also include an NFW halo \citep{NFW} to model the dark matter component of the potential. We find that the pulsar data is not able to constrain the scale length of the halo component; this is likely because the radial variations in the halo density are small compared to the region probed by the pulsar data. As a result, we fix the scale length to be $r_s$ = 15.6 kpc, which matches the scale length of the halo component of the \textit{Gala} \textbf{MilkyWayPotential2022} model \citep{Gala}. The pulsar data is able to constrain the mass of the NFW component, however, which is optimized along with the disk parameters. Different choices for the NFW scale length within reasonable limits did not dramatically change the halo mass of the model fits.

The optimized parameters for each disk model are provided in Table \ref{tab:model_params}. Each model consists of the disk component plus the halo component. Additionally, we allowed the center of mass of the halo component to shift vertically along the $z$ axis relative to the disk midplane (see Section \ref{sec:disk_halo_offset} for more detail on this). The fit values for the various scale lengths and heights, as well as the total masses of the disk and halo components, are broadly consistent with the scale lengths and heights for various models of the thin disk of the Galaxy \citep{BlandHawthornGerhard2016}. 

We also list Akaike Information Criterion \citep[AIC,][]{AIC} for each fit, as well as the \textit{Gala} \textbf{MilkyWayPotential2022} model. AIC is a way of quantifying goodness-of-fit that penalizes additional parameters in order to discourage overfitting, and is defined \begin{equation}
    \rm{AIC} = 2k - 2\mathcal{L}, 
\end{equation} where $k$ is the number of fit model parameters. The model fit with the lowest AIC is ostensibly the best choice of model for the data; note that only differences in AIC matter (the absolute value of AIC includes a constant that depends on the input data, which will be identical for every model) so we subtract a constant from all listed AICs to enable easier comparison. Our three disk models have essentially the same AIC, indicating that they are equally good fits (a $\Delta$AIC $>$ 8 is typically considered strong evidence against a specific model). The \textit{Gala} \textbf{MilkyWayPotential2022} and \textit{Galpy} \textbf{MWPotential2014} models, which are fit to a variety of kinematic observations, both have very high $\Delta$AIC. This implies that the acceleration data are not consistent with the modern kinematic models.

The main difference between our models and the kinematic models is the mass of the Galaxy. Our models predict a somewhat more massive disk and halo, with a disk mass of roughly 8$\times10^{10}$ M$_\odot$ and a halo mass of about 10$^{12}$ M$_\odot$. For comparison, the mass of the disk is 4.8$\times10^{10}$ M$_\odot$ in the \textit{Gala} \textbf{MilkyWayPotential2022} model and 6.8$\times10^{10}$ M$_\odot$ in the \textit{Galpy} \textbf{MWPotential2014} model; the halo mass inside the virial radius is roughly 9.4$\times10^{11}$ M$_\odot$ in the \textit{Gala} \textbf{MilkyWayPotential2022} model and 8.0$\times10^{11}$ M$_\odot$ in the \textit{Galpy} \textbf{MWPotential2014} model.

\subsection{Local Density Constraints}

\begin{figure}[]
    \centering
    \includegraphics[width=\linewidth]{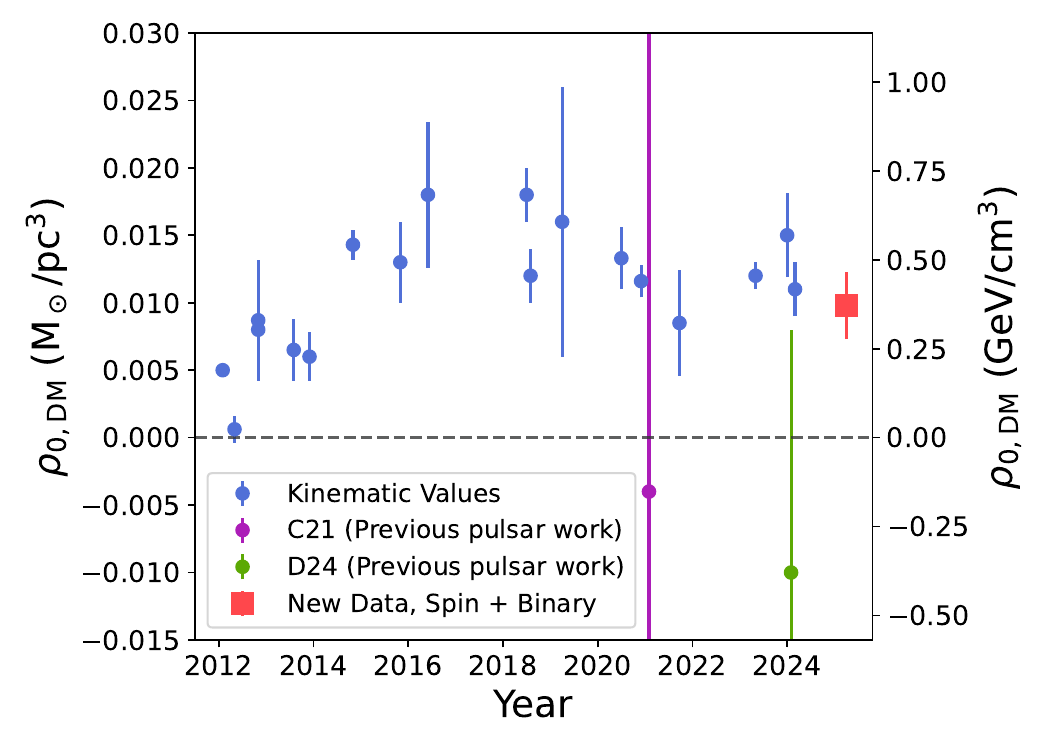}
    \caption{Various literature measurements of the dark matter density in the midplane ($\rho_\mathrm{0,DM}$). Using the expanded dataset and updated potential models (red square), we now obtain a value of $\rho_\mathrm{0,DM}$ that is similar to those obtained from kinematic data, and our error bars are significantly smaller than those of previous pulsar studies. }
    \label{fig:rho_dm}
\end{figure}

Our goal is to measure the local dark matter density in the mid-plane, $\rho_{0,\mathrm{DM}}$. Prior work using pulsar accelerations \citep{Chakrabarti2021,Donlon2024} converted from the total midplane density to the mid-plane dark matter density by subtracting the density of stars and gas in the mid-plane, also known as a ``baryon budget''. These baryon budgets typically have statistical uncertainties of about 5--10\%. However, a comparison of the baryon budgets from different studies shows that these values can vary between studies by several times the reported statistical uncertainties \citep[for example, see ][]{Bienayme2014,McKee2015,Bovy2017,Lim2023}, which hints at a significant unreported systematic uncertainty in the kinematically-derived baryon budgets.

Here, we adopt a different and more self-consistent approach.  Each of the disk model fits produces a value for the density in the midplane (the Oort limit), \begin{equation}
    \rho_0 = \rho_{0,\mathrm{bary}} + \rho_{0,\mathrm{DM}},
\end{equation} which consists of the midplane baryonic density $\rho_{0,\mathrm{bary}}$ (taken to be the density of the disk model) and the midplane dark matter density $\rho_{0,\mathrm{DM}}$ (taken to be the density of the NFW halo component). This allows us to obtain a value for the dark matter density independent of an external baryon budget (which have uncertainties much larger than our statistical uncertainties on $\rho_0$ and $\rho_{0,\mathrm{DM}}$). 

Across all fits, the value of the Oort Limit is $\rho_0$ = 0.108 $\pm$ 0.008 \textit{stat}. $\pm$ 0.011 \textit{sys} M$_\odot$/pc$^3$, and the value of the dark matter density is found to be $\rho_{0,\mathrm{DM}}$ = 0.0098 $\pm$ 0.0025 \textit{stat.} $\pm$ 0.0003 \textit{sys}. M$_\odot$/pc$^3$, which is equal to 0.37 $\pm$ 0.10 GeV/cm$^{3}$. This is by far the most precise measurement of the Oort Limit and the dark matter midplane density made using direct acceleration measurements to-date. By considering only the disk component of the model fits, we obtain a value for the baryon budget of $\rho_{0,\mathrm{b}} = 0.097 \pm 0.017 $ M$_\odot$/pc$^3$.

\subsection{Kinematic vs. Acceleration Studies}

Figure \ref{fig:rho_dm} shows a comparison of various literature values for $\rho_{0,\mathrm{DM}}$ and the recent values obtained using pulsar direct acceleration measurements (references and data can be found in Appendix A of \citealt{Donlon2024}). Previously, the results from pulsar measurements were much lower than kinematic estimates of $\rho_{0,\mathrm{DM}}$ and had large uncertainties, in part due to their use of $\alpha$ models for the potential and external baryon budgets. Our new measurement has much smaller uncertainty compared to previous pulsar-based measurements, and is in good agreement with the kinematic measurements of $\rho_{0,\mathrm{DM}}$. 

It should be noted that the kinematic estimates of the Oort limit (and therefore the local dark matter density) are likely biased high. This is because disequilibrium effects, such as variations in surface density across the disk, can lead to an overestimated average determination of the Oort limit \citep{HainesDonghia2019}. The values for the dark matter density obtained by these studies will be biased unless they thoroughly compensate for these disequilibrium effects. As a result, there is a significant systematic uncertainty in the kinematic values of the local dark matter density that can make the true uncertainties of these values larger than what is reported.

\subsection{Discussion}


We have analyzed the MW acceleration profile using equilibrium disk potentials -- that is, models which presume that the potential of the MW is azimuthally symmetric, and symmetric above and below the midplane -- plus perturbations due to various disequilibrium effects. This analysis relies on parameterized models with fixed shapes for the MW potential. As a result, the methods used here may not be flexible enough to capture the large number of disequilibrium accelerations in the Solar neighborhood \citep{Donlon2024}, which could cause increased systematic uncertainties and/or biases when measuring the local density. 

This emphasizes the importance of developing approaches which utilize accelerations in a non-parametric way -- in essence, methods that do not assume any specific distributions or shapes for the Galactic potential. Such studies might have more success in measuring local perturbations to the acceleration field (and therefore the density field) than works that fit parameterized potential models to observed accelerations. 

Perhaps most importantly, the systematic biases of Oort limit measurements using pulsar accelerations are still not well understood; further work needs to be done to characterize how pulsar acceleration measurements compare to other methods of estimating the Oort limit and density asymmetries.  Upcoming eclipse timing measurements of the Galactic acceleration \citep{Chakrabarti2022} should provide a comparative standard.

Although we assumed that the disk component of the potential is purely baryonic, some theories of dark matter have been postulated that would result in a thin disk of dark matter in the Galactic midplane (see Section 7.2 of \citealt{McKee2015} for a review of this idea). A thin disk of dark matter would cause our estimate for the local dark matter density to be incorrect. However, such a structure has been disfavored by the latest kinematic data \citep{Schutz2018}, and faces significant theoretical issues -- including the gravitational stability of a very thin dark matter disk \citep{McKee2015}, as well as the properties required for dark matter to form such a structure \citep{Fan2013}. Due to these issues we choose not to consider a dark matter disk in our models, although at this time the direct acceleration measurements cannot rule out the existence of a dark thin disk. 


\section{Conclusions} \label{sec:conclusion}

Direct acceleration measurements from pulsar timing data contain a wealth of information that is immediately relevant for studies of the structure of our Galaxy. Until now, acceleration studies were limited to either aggregate constraints on acceleration from spins, such as Phillips et al. \cite{Phillips2021}, or measurements of the accelerations of individual sources for only binary MSPs \citep{Chakrabarti2021,Moran2023,Donlon2024}. 

For the first time, we create a procedure for measuring accelerations on a source-by-source basis using only the spin information of MSPs. We show that the spindown rate due to magnetic braking and a pulsar's estimated minimum surface magnetic field strength are directly proportional, and provide a relation for estimating the intrinsic spindown of a pulsar given only its spin period and time derivative of the spin period. However, this relationship only holds true for MSPs with $B_{\rm Surf}<3\times10^8$ G and characteristic age $\tau>5$ Gyr. For the pulsars that satisfy these constraints, the empirical model is able to reliably estimate the acceleration of each pulsar to roughly the same level of precision as the acceleration measurement from binary orbital period information. 

There are 26 MSPs that satisfy the \bsurf and characteristic age constraints and have spin period information, but no binary orbital period information. The addition of these 26 new sources effectively doubles the number of available pulsar measurements -- which, when combined with the 27 existing binary MSP measurements, results in a total of 53 datapoints. By only requiring spindown data for a given pulsar, this also opens up the possibility of utilizing X-ray and $\gamma$-ray timing of pulsars to obtain accelerations \citep{Deneva2019,Deneva2021,Zheng2024}, which was previously not possible given the difficulty of obtaining precise orbital period information for binary millisecond pulsars at these wavelengths. This development showcases the recent and rapid increase in available direct acceleration data that is expected to continue into the near future. 

The pulsar acceleration data contains a substantial asymmetry in the vertical acceleration profile, which was previously measured by Donlon et al. \cite{Donlon2024}. We point out that this local gradient in the vertical acceleration profile can be caused by at least two processes; the north-south density asymmetry in the disk star counts, and the offset of the MW halo and disk centers of mass. The combination of these two effects produces a vertical acceleration gradient that has the same shape and magnitude as the observed gradient. 

The expanded pulsar dataset allows us to obtain an updated measurement of the total density in the Galactic midplane, which we find to be $\rho_0$ = 0.108 $\pm$ 0.008 \textit{stat}. $\pm$ 0.011 \textit{sys} M$_\odot$/pc$^3$, and an updated measurement of the local dark matter density, which we calculate to be $\rho_{0,\mathrm{DM}}$ = 0.0098 $\pm$ 0.0025 \textit{stat.} $\pm$ 0.0003 \textit{sys}. M$_\odot$/pc$^3$. The uncertainty on this value is much smaller than those of previous works that utilize accelerations from pulsar timing data. This represents the first $>3\sigma$ measurement of the local dark matter density from direct acceleration measurements. Although previous pulsar acceleration studies produced very small values for the midplane dark matter density, our updated measurement is in good agreement with existing measurements of $\rho_{0,\mathrm{DM}}$.

\acknowledgments

We would like to thank Michael T. Lam, Alice Quillen, and Thomas Tauris for helpful conversations and ideas.  

Sukanya Chakrabarti acknowledges support from NASA EPSCoR CAN AL-80NSSC24M0104 and STSCI GO 17505.
Lawrence Widrow was supported by a Discovery Grant with the Natural Sciences and Engineering Research Council of Canada.

This work makes use of data from the Australia Telescope National Facility (ATNF) Pulsar Catalogue, which can be found at \url{ http://www.atnf.csiro.au/research/pulsar/psrcat}. 



\bibliographystyle{apsrev4-2}
\bibliography{main.bib}

\appendix

\section{Validity of the $B_\mathrm{Surf}$ Approximation} \label{app:bsurf_approx}

\begin{figure}
    \centering
    \includegraphics[width=\linewidth]{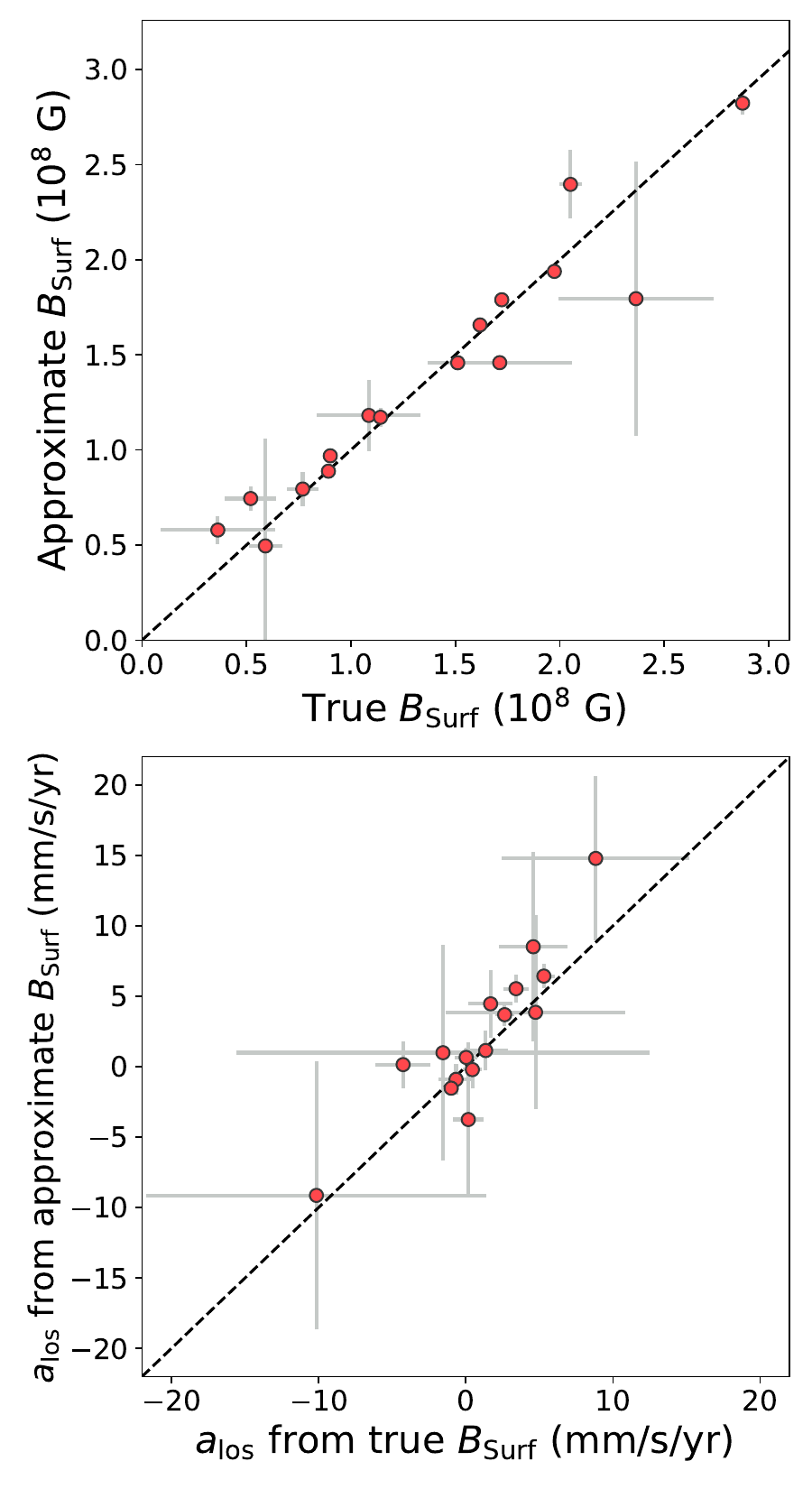}
    \caption{Comparison of the true value of \bsurf from Equation \ref{eq:bsurf_true} and the approximate value of \bsurf from Equation \ref{eq:bsurf_approx}. The dashed black line shows where the two values are equal. The approximation leads to a small amount of scatter in the inferred value of \bsurf -- and as a result, a small amount of scatter in the inferred acceleration from the empirical model.  }
    \label{fig:bsurf_approx}
\end{figure}

In Section \ref{sec:magnetic_braking_model}, we make an approximation of \bsurf in Equations \ref{eq:bsurf_true} and \ref{eq:bsurf_approx}. This is done so that we do not have to assume or measure a value of $\dot{P}_s^\mathrm{Gal}$ for each pulsar in our sample. However, this leads to a small amount of error in our value of \bsurf that is used in the empirical magnetic spindown model. 

Figure \ref{fig:bsurf_approx} shows the error that arises from this approximation for the D24 dataset, which are pulsars for which we know the true value of $\dot{P}_s^{B}$ and therefore \bsurf from their binary orbital period information. The approximation leads to a small amount of scatter in the value of $B_\mathrm{Surf}$, which correspondingly leads to scatter in the inferred acceleration of these pulsars. This scatter appears to be of the same scale as the scatter in the true and model accelerations shown in the bottom left panel of Figure \ref{fig:fit_and_errors}, although the individual errors in \bsurf and the line-of-sight accelerations do not appear to be directly related. 

We conclude that our approximation in \bsurf is acceptable for these pulsars, although it is probably a significant source of the error in the modeled accelerations of these pulsars.

\section{Adding an Intrinsic Scatter Term to the Linear Model} \label{app:int_err}

\begin{figure}
    \centering
    \includegraphics[width=\linewidth]{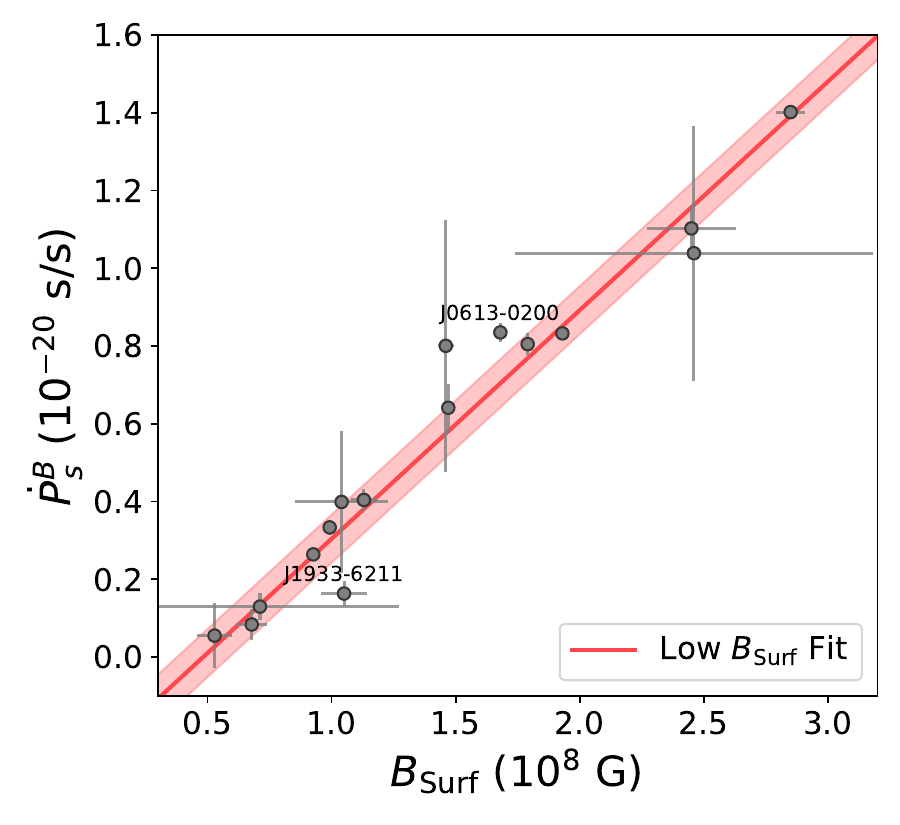}
    \caption{Linear fit to the low-\bsurf pulsar data, including an intrinsic scatter term (shown as the shaded red region).}
    \label{fig:int_err}
\end{figure}

It is clear based on the top-left panel of Figure \ref{fig:fit_and_errors} that there is significant statistical noise in the data that is not captured by the fit. One way of quantifying this is by introducing an intrinsic error (scatter) term, $\sigma_\mathrm{int}$, which represents the $\dot{P}_s^B$ of each point being randomly sampled according to a normal distribution with a standard deviation of $\sigma_\mathrm{int}$. 

This makes the corresponding log-likelihood function: \begin{align}
    -\mathcal{L} = \frac{1}{2}\sum_i \Bigg[ \log\left( 2 \pi (\sigma_i^2 + \sigma_\mathrm{int}^2)\right) \\ \nonumber 
     + \frac{\left( \dot{P}_{s,i,\mathrm{obs}}^B - \dot{P}_{s,i,\mathrm{model}}^B \right)^2}{\sigma_i^2 + \sigma_\mathrm{int}^2} \Bigg]. 
\end{align} 

When including this intrinsic error term, new fit parameters for the linear model remain essentially unchanged compared to the fit without the intrinsic error term. The fit intrinsic error term is $\sigma_\mathrm{int} = (6.2\pm0.7)\times10^{-22}$ s/s; this is shown in figure \ref{fig:int_err} as the red region around the line of best fit. Most of the pulsars lie within this band (particularly if one considers their error bars), although there are a couple sources that do not fall within this region -- these sources have been labeled for convenience.

\end{document}